\documentclass[12pt]{amsart}
\usepackage{amsmath}
\usepackage{amsxtra}
\usepackage{amscd}
\usepackage{amsthm}
\usepackage{amsfonts}
\usepackage{amssymb}
\usepackage{eucal}
\def\({\left(}
\def\){\right)}
\newcommand{\alb}{\mbox{\boldmath$\al$}}
\newcommand{\kapb}{\mbox{\boldmath$\kappa$}}
\newcommand{\scalb}{\mbox{\boldmath$\scriptstyle\al$}}
\newcommand{\taub}{\mbox{\boldmath$\tau$}}

\newcommand{\ket}[1]{{| #1 \rangle}}      

\newcommand{\cb}{\mathbf{c}}
\newcommand{\bb}{\mathbf{b}}
\newcommand{\ab}{\mathbf{a}}
\newcommand{\tb}{\mathbf{t}}
\newcommand{\jb}{\mathbf{j}}
\renewcommand{\sb}{\mathbf{s}}

\newcommand{\nn}{\nonumber}
\newcommand{\bea}{\begin{eqnarray}}
\newcommand{\ena}{\end{eqnarray}}
\def\bel{\begin{eqnarray}}
\def\enl{\end{eqnarray}}
\newcommand{\be}{\begin{eqnarray*}}
\newcommand{\en}{\end{eqnarray*}}
\newcommand{\ba}{\begin{array}}
\newcommand{\ea}{\end{array}}

\newcommand{\C}{{\mathbb C}}
\newcommand{\Z}{{\mathbb Z}}

\newcommand{\cA}{\mathcal{A}}
\newcommand{\cP}{\mathcal{P}}

\newcommand{\cC}{\mathcal{C}}
\newcommand{\cD}{\mathcal{D}}

\newcommand{\cV}{\mathcal{V}}

\newcommand{\res}{{\rm res}}
\newcommand{\id}{{\rm id}}

\newcommand{\tr}{{\rm tr}}

\newcommand{\Tr}{{\rm Tr}}

\newcommand{\End}{\mathop{\rm End}}

\newenvironment{tenumerate}{
  \begin{enumerate}
  
  }{\end{enumerate}}
\newcommand{\bi}{\begin{tenumerate}}
\newcommand{\ei}{\end{tenumerate}}
\newcommand{\isoto}[1][]%
{{\mathop{\buildrel{\sim}\over\longrightarrow}\limits_{#1}}}

\def\[{\left[}
\def\]{\right]}

\newcommand{\g}{\gamma}
\newcommand{\al}{\alpha}
\newcommand{\e}{\epsilon}

\newcommand{\s}{\sigma}
\newcommand{\z}{\zeta}

\newcommand{\psit}{\tilde{\psi}}
\newcommand{\psih}{\hat{\psi}}

\newcommand{\Psih}{\widehat{\Psi}}
\newcommand{\Phih}{\widehat{\Phi}}

\numberwithin{equation}{section}
\newtheorem{thm}{Theorem}[section]
\newtheorem{prop}[thm]{Proposition}
\newtheorem{lem}[thm]{Lemma}

\newtheorem{cor}[thm]{Corollary}

\newcommand{\cL}{\mathcal{L}}
\newcommand{\cLh}{\widehat{\mathcal{L}}}
\newcommand{\cW}{\mathcal{W}}
\newcommand{\bS}{\mathbb{S}}
\newcommand{\bJ}{\mathbb{J}}
\newcommand{\bR}{\mathbb{R}}
\newcommand{\bQ}{\mathbf{Q}}

\newcommand{\ao}{\mathbf{a}}
\newcommand{\Ob}{\mbox{\boldmath$\Omega $}}

\newcommand{\bk}{\mathbf{k}}

\begin{document} 

\begin{title}[Fermionic basis]
{Fermionic basis for space of operators\\
 in the XXZ model}
\end{title}
\date{\today}
\author{H.~Boos, M.~Jimbo, T.~Miwa, F.~Smirnov and Y.~Takeyama}
\address{HB: Physics Department, University of Wuppertal, D-42097,
Wuppertal, Germany\footnote{
on leave of absence from 
Skobeltsyn Institute of Nuclear Physics, 
MSU, 119992, Moscow, Russia
}}\email{boos@physik.uni-wuppertal.de}
\address{MJ: Graduate School of Mathematical Sciences, The
University of Tokyo, Tokyo 153-8914, Japan}\email{jimbomic@ms.u-tokyo.ac.jp}
\address{TM: Department of Mathematics, Graduate School of Science,
Kyoto University, Kyoto 606-8502, 
Japan}\email{tetsuji@math.kyoto-u.ac.jp}
\address{FS\footnote{Membre du CNRS}: Laboratoire de Physique Th{\'e}orique et
Hautes Energies, Universit{\'e} Pierre et Marie Curie,
Tour 16 1$^{\rm er}$ {\'e}tage, 4 Place Jussieu
75252 Paris Cedex 05, France}\email{smirnov@lpthe.jussieu.fr}
\address{YT:  Institute of Mathematics, 
Graduate School of Pure and Applied Sciences, Tsukuba University, 
Tsukuba, Ibaraki 305-8571, Japan}
\email{takeyama@math.tsukuba.ac.jp}

\begin{abstract}
{In the recent study of correlation functions for the infinite 
XXZ spin chain, a new pair of anti-commuting operators $\bb(\z),\cb(\z)$
was introduced. They act on the space of {\it quasi-local} operators,
which are 
local operators multiplied by the disorder operator $q^{2\al S(0)}$
where $S(0)=\frac12\sum_{j=-\infty}^0\s^3_j$.
For the inhomogeneous chain with the spectral
parameters $\xi_k$, these operators have simple poles at $\z^2=\xi_k^2$.
The residues are denoted by $\bb_k,\cb_k$. At $q=i$,
we show that the operators $\bb_k,\cb_k$ are cubic monomials in
free fermions. In other words, the action of these operators is very simple
in the fermion basis. We give an explicit construction of these fermions.
Then, we show that the existence of the fermionic basis
is a consequence of the Grassmann relation, the equivariance with respect to
the action of the symmetric group and the reduction property,
which are all valid for the operators $\bb_k,\cb_k$ in the case of 
generic $q$.}
\end{abstract}
\maketitle
\bigskip

\section{Introduction}
This is a continuation of our previous paper \cite{BJMST}
in which we studied the infinite XXZ spin chain
with the Hamiltonian
\be
H_{XXZ}=\frac12\sum_{k=-\infty}^\infty\left(\s^1_k\s^1_{k+1}
+\s^2_k\s^2_{k+1}+\Delta\s^3_k\s^3_{k+1}\right),\quad
\Delta=\frac{q+q^{-1}}2.
\en
Before going into the details of the present paper, we review
the main points of \cite{BJMST}.
In that paper we gave an algebraic formula for the
vacuum expectation values 
of quasi-local operators of the form
\bea
q^{\alpha\sum_{k=-\infty}^0\s^3_k}\mathcal O\label{QLO}.
\ena
Here, $\mathcal O$ is a local operator, and $\al$ is a disorder parameter.
We denote by
\be
S(j)=\frac12\sum_{k=-\infty}^j\s^3_k
\en
the total spin operator on the left half of the chain.

Let $\cW_\al$ be the space of quasi-local operators of this form.
The main ingredient in our formula was a pair of operators
$\bb(\z,\al),\cb(\z,\al)$ acting on the spaces of quasi-local operators,
\be
\bb(\z,\al):\cW_\al\rightarrow\cW_{\al+1},\
\cb(\z,\al):\cW_\al\rightarrow\cW_{\al-1}.
\en
It is convenient to introduce the space
\be
\cW_{[\al]}=\oplus_{k\in\Z}\cW_{\al+k},
\en
and use the notation $\bb(\z),\cb(\z):\cW_{[\al]}\rightarrow\cW_{[\al]}$.

The vacuum expectation valued are given by the formula
\be
\frac{\langle{\rm vac}|
q^{2\alpha S(0)}\mathcal O|{\rm vac}\rangle}
{\langle{\rm vac}|q^{2\alpha S(0)}|{\rm vac}\rangle}
=\mathbf{tr}^{\al}\(
e^{\mbox{\scriptsize\boldmath{$\Omega$}}}\(q^{2\al S(0)}\mathcal{O}\)\)\,, 
\en
where $\mathbf{tr}^{\al}$ stands for the weighted trace 
\be
&&\mathbf{tr}^{\al}(X)=\cdots\tr _1^{\al}\ \tr _2^{\al}\ 
\tr _3^{\al}\cdots (X)\quad (X\in\cW_\al),\\
&&\text{tr}^{\alpha}(x)=
\frac 1 {q^{\frac {\al }2}+q^{-\frac {\al }2}}
\text{tr} \(q^{-\frac 1 2 \al \sigma ^3}x\)
\quad (x\in \End(\C^2))\,,
\en
and on each $\cW_\al$, $\Ob$ admits an expression 
\bea
&&\Ob=
 -\res_{\z_1,\z_2=1}\left(\omega(\z_1/\z_2,\al)\bb(\z_1)\cb(\z_2)
\frac{d\z_1}{\z_1}\frac{d\z_2}{\z_2}\right)\,.
\label{Omega}
\ena
Here $\omega(\z,\al)$ is a known scalar function
(whose explicit formula can be found in (1.5), \cite{BJMST}).

The operators $\bb(\z),\cb(\z)$ are obtained from
the operators $\bb_{[k,l]}(\z),\cb_{[k,l]}(\z)$
which act on the direct sum of the finite dimensional spaces
\be
\cV_{[\al],[k,l]}=\oplus_{j\in\Z}\cV_{\al+j,[k,l]},\
\cV_{\al+j,[k,l]}\simeq\End\left((\C^2)^{\otimes(l-k+1)}\right)
\en
by taking an inductive limit when $k\rightarrow-\infty,l\rightarrow\infty$.
Here the index $[k,l]$ indicates
the finite interval $[k,l]\subset\Z$ of the spin chain.

We use $\cb^+=\cb$ and $\cb^-=\bb$ in the sequel. 
On each finite interval $[k,l]$, 
the operator $\cb^\pm_{[k,l]}(\z)$ has the form
\be
\cb^\pm_{[k,l]}(\z)=-\frac{1}{2}
\sum_{j=1}^{l-k+1}\frac{\cb^{\pm(j)}_{[k,l]}}{(\z-1)^j}\,.
\en

The operators $\bb^{(j)}_{[k,l]},\cb^{(j)}_{[k,l]}$ depend 
on $k,l$ in such a way that the reduction properties hold:
\be
&&\cb^\pm_{[k,l]}(X_{[k,l-1]})=\cb^\pm_{[k,l-1]}(X_{[k,l-1]})\,,\\
&&\cb^\pm_{[k,l]}(q^{\al\s^3_k}X_{[k+1,l]})=q^{(\al\mp1)\s^3_k}
\cb^\pm_{[k+1,l]}(X_{[k+1,l]})\,.
\en
Here $X_{[k,l]}$ denotes an element of $\mathcal V_{\al,[k,l]}$.

In \cite{BJMST}, the operator $\cb^\pm_{[k,l]}(\z)$ was constructed
from the trace of a monodromy matrix whose auxiliary space consists
of the tensor product of $\C^2$ and the $q$ oscillator representations
$W^\pm$. This is similar to the construction of $Q$ operators in \cite{BLZ}.
However our construction differs from that paper in two respects. 
We take the trace of the off-diagonal entry of the monodromy matrix which is
triangular with respect to the auxiliary space $\C^2$;
and we consider only the singular part in the expansion at $\z=1$, since
otherwise the operator is dependent on the way of triangularization.
As a spin-off of our construction we obtained the Grassmann relation
among the operators $\cb^\pm(\z)$:
\be
\cb^{\varepsilon_1}(\z_1)\cb^{\varepsilon_2}(\z_2)=-
\cb^{\varepsilon_2}(\z_2)\cb^{\varepsilon_1}(\z_1)\,.
\en

In this paper, we further investigate the structure of the operators
$\cb^\pm(\z)$. We consider the inhomogeneous lattice with
the spectral parameters $\xi_k$ $(k\in\Z)$. There are two advantages
in considering the inhomogeneous case. The singularities of the trace which lie
at $\z^2=1$ in the homogeneous case split into simple poles at $\z^2=\xi_k^2$
in the inhomogeneous case. Therefore the operator $\cb^\pm_{[k,l]}(\z)$
is decomposed into a partial fraction
\be
\cb^\pm_{[k,l]}(\z)=-\frac{1}{2}
\sum_{j=k}^l\frac{\xi_j\cb^\pm_{j,[k,l]}}{\z-\xi_j}.
\en
This is the first point.
The reduction property and the Grassmann relation persist in the inhomogeneous
case. Passing to the inductive limit, we obtain
a set of operators $\cb^\pm_j$ acting on $\cW_{[\al]}$
and satisfying the Grassmann relation:
\be
\cb^{\varepsilon_1}_{j_1}\cb^{\varepsilon_2}_{j_2}=-
\cb^{\varepsilon_2}_{j_2}\cb^{\varepsilon_1}_{j_1}.
\en
Moreover, in the inhomogeneous case, the operators $\cb^\pm_j$
enjoy the equivariance with respect to the action of the 
infinite symmetric group inherited from the $R$ matrix symmetry:
\be
\sb_i\cdot\cb^\pm_j=\cb^\pm_{s_i(j)}\cdot\sb_i,\quad
\sb_i\in\mathfrak S_\infty.
\en
This is the second point.

Although the operators $\cb^\pm_j$ satisfy the Grassmann relation 
and annihilate some elements in $\cW_{[\al]}$,
they do not constitute the annihilation part of the free fermion algebra.
We see this because
they have a large kernel in common, which is spanned by the
$\mathfrak S_\infty$ orbits of $q^{2\al S(j)}$ $(j\in\Z)$.
We call elements in the common kernel
{\it vacuum states}. The existence of a large kernel is a simple consequence of the
reduction property and the equivariance with respect to 
the $\mathfrak S_\infty$ action.
In fact, the reduction property holds in a stronger sense:
\be
\cb^\pm_{l,[k,l]}\bigl(V_{\al,[k,l]}\bigr)\subset
\mathcal V_{\al,[k,l-1]}\otimes{\rm id}\,.
\en
Here ${\rm id}$ is the identity operator on the $l$-th tensor component.
This implies that there exists an $\mathfrak S_\infty$-
equivariant filtration of the space $\cW_{[\al]}$, 
\be
&&F^0\cW_{[\al]}\subset\cdots\subset F^n\cW_{[\al]}\subset F^{n+1}\cW_{[\al]}
\subset\cdots\cW_{[\al]}\,,\\
&&\sb_i\left(F^n\cW_{[\al]}\right)\subset F^n\cW_{[\al]}\,,
\en
such that $F^0\cW_{[\al]}$ is the space of the vacuum states and
the operator $\cb^\pm_j$ decreases the `particle' number $n$:
\be
\cb^\pm_j(F^n\cW_{[\al]})\subset F^{n-1}\cW_{[\al]}\,.
\en

In the case $q=i$, it is well-known that the Hamiltonian
is diagonalized by the Jordan-Wigner transformation.
The latter turns the tensor product $(\C^2)^{\otimes n}$ 
into the irreducible representation of the free fermion algebra
with $2n$ generators 
\be
\psi^\pm_j=\s^\pm_ji^{\mp\sum_{k=1}^{j-1}\s^3_k}\quad (1\leq j\leq n).
\en
We introduce a fermion algebra with $4n$ generators 
$\widehat{\Psi}^\pm_j,\widehat{\Phi}^\pm_j$ $(1\leq j\leq n)$,
\be
[\widehat{\Psi}^{\varepsilon_1}_{j_1},
\widehat{\Psi}^{\varepsilon_2}_{j_2}]=0,\
[\widehat{\Phi}^{\varepsilon_1}_{j_1},
\widehat{\Phi}^{\varepsilon_2}_{j_2}]=0,\
[\widehat{\Psi}^{\varepsilon_1}_{j_1},
\widehat{\Phi}^{\varepsilon_2}_{j_2}]=
\delta_{\varepsilon_1,-\varepsilon_2}\delta_{j_1,j_2}\,,
\en
which act on $\End\left((\C^2)^{\otimes n}\right)$ irreducibly.
They are linear combinations of the left and the right multiplications 
by $\psi^\pm_j$, equivariant relative to the $\mathfrak S_n$ action, 
and satisfy the reduction property which enables us to extend their definition
to the infinite lattice. It is convenient to introduce the fermion number
operators $N^\pm_k=\widehat{\Phi}^\pm_k\widehat{\Psi}^\mp_k$. The vacuum states
$\C q^{2\al S(j)}$ $(j\in\Z)$ are characterized by the fermion numbers
\be
N^\pm_k=\begin{cases}
1&\hbox{ if $k\leq j$;}\\
0&\hbox{ if $k\geq j+1$.}
\end{cases}
\en
Finally, we modify these fermions to another set of free fermions
$\chi^\pm_k,\chi^{*\pm}_k$ satisfying $N^\pm=\chi^{*\pm}_k\chi^{\mp}_k$, 
and obtain the simple formula
\be
\cb^\pm_k=\chi^\pm_k\chi^\mp_k\chi^{*\pm}_k.
\en
Namely, the operator $\cb^\pm_k$ is a fermion annihilation operator: if
the fermion number of a state is $N^\mp_k=1$, it is changed to
$N^\mp_k=0$; otherwise the state is annihilated.

Let us return to generic $q$.
The properties of the operators $\cb^\pm_k$ formulated above 
are enough to deduce the existence of a fermionic basis of the space
$\cW_{[\al]}$ on which they act exactly as in the $q=i$ case
(see Theorem \ref{BASIS}). This is our conclusion.
Such a fermionic basis is by no means unique.
We hope to discuss a more explicit construction in a future publication.

The plan of this paper is as follows. In Section 2 we recall the definition
of $\bb(\z)$ and $\cb(\z)$ given in \cite{BJMST}. Appendix A is a supplement
to this section, giving a proof of some operator identities.
Section 3 with the detailed calculations in Appendix B gives the complete
details of the $q=i$ case. Section 4 is devoted to the existence proof of the
fermionic basis for generic $q$.

\section{Operators $\bb$ and $\cb$}\label{sec:2}

In this section, we give the definition of the operators
$\bb(\z)$, $\cb(\z)$ and discuss their basic properties: 
the Grassmann relation \eqref{GRASS}, the equivariance
with respect to the symmetric group \eqref{INV}, and
the reduction properties \eqref{LREDUC} and \eqref{RREDUC}.

Throughout the paper, $q$ denotes a non-zero complex number 
and $V$ a two-dimensional vector space with fixed basis $v_+, v_-$. 

\subsection{Quantum spaces}
We consider an inhomogeneous spin chain with spectral parameters 
$\xi_j$ $(j\in\Z)$. 
They are attached to the `quantum' spaces $V_j$ ($j\in\Z$),  
where $V_j$ is a copy of $V$. 

In this and the next section, we fix a positive integer $n$ and 
consider a finite segment $[1,n]$ of the infinite spin chain. 
We work with the space 
\be
\cV=\End(V_1\otimes\cdots\otimes V_n)
\en
consisting of operators which 
act non-trivially on the lattice sites $j=1,\cdots,n$. 
We denote by $P_{i,j}\in\End(\mathcal V)$ 
the transposition of the $i$-th and the $j$-th tensor components. 
We consider the transfer matrix acting on the space $\mathcal V$. 
It depends on a parameter $\al$ playing the role of a boundary condition.  
Later we shall extend $\cV$ to an inductive limit,  
where the index $j$ runs over $\Z$.
The parameter $\alpha$ enters the construction of this  
inductive limit as well.  

To make distinction, we call an element of $\mathcal V$ a {\it state}, 
and an element of $\End(\mathcal V)$ an {\it operator}.  
When it is necessary,
we extend the coefficient field of $\mathcal V$ or $\End(\mathcal V)$
to the field of rational functions in the variables 
$q^\al,\xi_1,\ldots,\xi_n$.
We denote by $r_{i,j}$ the operation of exchanging 
spectral parameters $\xi_i\leftrightarrow\xi_j$. 

The total spin operator $S=\frac12\sum_{j=1}^n\s^3_j$ belongs to $\mathcal V$,
and its adjoint action
\be
\bS(X)=[S,X]\ \quad (X\in\mathcal V)
\en
belongs to $\End(\mathcal V)$.
The space $\mathcal V$ decomposes as
$\mathcal V=\bigoplus_{s\in\Z}\mathcal V^{(s)}$ where
$\mathcal V^{(s)}=\{X \in\mathcal V\,|\,\bS(X)=sX\}$.

The spin reversal $\bJ$ is an operator acting on $\mathcal V$
\be
\bJ(X)= \prod_{j=1}^n\s^1_j\cdot X\cdot \prod_{j=1}^n\s^1_j\,.
\en
When we extend the coefficient field, we use the operator
$\jb$ given by
\be
\jb(X)=\bJ(X)|_{q^\al\rightarrow q^{-\al}}\,.
\en

\subsection{$R$ matrix}
The space $V$ is endowed with a structure of a two-dimensional 
evaluation module of the quantum affine 
algebra $U_q(\widehat{\mathfrak{sl}}_2)$. 
We will need only the formula for the associated $R$ matrix:
\be
R(\z)=(q\z-q^{-1}\z^{-1})\cdot
\begin{pmatrix}1
\\&\beta(\z)&\g(\z)\\&\g(\z)&\beta(\z)\\&&&1
\end{pmatrix}\,,
\en
where 
\be
\beta(\z)=\frac{(1-\z^2)q}{1-q^2\z^2},
\quad
\g(\z)=\frac{(1-q^2)\z}{1-q^2\z^2}\,.
\en
The matrix $R(\z)$ is a Laurent polynomial, while
$R(\z)^{-1}$ has poles at $\z^2=q^{\pm2}$.

The $R$ matrix gives an action of the symmetric group $\mathfrak S_n$ on
$\mathcal V$, the simple reflection $s_i$ being represented by
\be
\sb_i=r_{i,i+1}\check \bR_{i,i+1}(\xi_i/\xi_{i+1}). 
\en
Here
\be
&&\bR_{i,i+1}(\xi_i/\xi_{i+1})(X)=\check R_{i,i+1}(\xi_i/\xi_{i+1})X
\check R_{i,i+1}(\xi_i/\xi_{i+1})^{-1},\\
&&\check R_{i,i+1}(\z)=P_{i,i+1}R_{i,i+1}(\z).
\en

We use the monodromy matrix 
\be
T_a(\z)=R_{a,n}(\z/\xi_n)\cdots R_{a,1}(\z/\xi_1).
\en
Here and after, the suffix $a$ indicates the two-dimensional 
auxiliary space $V_a\simeq V$.

The total spin operator which commutes with $R_{a,j}(\z)$ is given by
$\frac12\s^3_a+S$, 
\be
[{\textstyle\frac12}\s^3_a+S,R_{a,j}(\z)]=0.
\en
The transfer matrix $\tb(\z,\al)$
acting on $\mathcal V$ is defined by using the inverse
adjoint action of the monodromy matrix.
\be
\tb(\z,\al)(X)=\tr_a(q^{-\al\s^3_a}T_a(\z)^{-1}XT_a(\z))\quad (X\in\mathcal V).
\en
Here $\tr_a=\tr_{V_a}$ stands for the trace over $V_a$, which gives a functional
\be
\tr_a(q^{-\al\s^3_a}\,\cdot\,)\,:\,\End(V_a)\rightarrow\C q^\al\oplus\C q^{-\al}.
\en
We have
\be
\sb_i\cdot \tb(\z,\al)=\tb(\z,\al)\cdot\sb_i,\
\jb\cdot \tb(\z,\al)=\tb(\z,\al)\cdot\jb.
\en

\subsection{Oscillator algebra}
{}Following \cite{BLZ}, we define another kind of monodromy matrices.  
They have the $q$ oscillator algebra $Osc$ as auxiliary spaces.
The oscillator algebra is 
generated by $\ao,\ao^*,q^{\pm D}$ with the relations
\be
q^D\ab q^{-D}=q^{-1}\ab,\
q^D\ab^*q^{-D}=q\ab^*,\
\ab\ab^*=1-q^{2D+2},\
\ab^*\ab=1-q^{2D}.
\en
We use the suffix $A$ indicating the infinite dimensional auxiliary space
$Osc_A$ and its generators $\ab_A$, $\ab^*_A$, $D_A\in Osc_A$.

We shall use two representations $W^\pm$ of $Osc$, 
\be
W^+=\oplus_{k\geq0}\C|k\rangle,\
W^-=\oplus_{k\leq-1}\C|k\rangle.
\en
The action is
\be
q^D|k\rangle=q^k|k\rangle,\
\ab|k\rangle=(1-q^{2k})|k-1\rangle,\
\ab^*|k\rangle=(1-\delta_{k,-1})|k+1\rangle.
\en
The $L$ operator $L^\pm(\z)$ belongs to $Osc\otimes\End(V)$.
We have
\be
&&L^+(\z)
=i\z^{-1/2}q^{-1/4}
(1-\z\ab^*\s^+-\z\ab\s^--\z^2q^{2D+2}\tau^-)q^{\s^3D},\\
&&L^-(\z)=\s^1L^+(\z)\s^1.
\en
Here $\tau^\pm=\s^\pm\s^\mp$. 
The inverse of $L^+(\z)$ is given by
\be
&&L^+(\z)^{-1}=\frac1{\z-\z^{-1}}\overline L^+(\z),\\
&&
\overline L^+(\z)=
i\z^{-1/2}q^{1/4}\cdot
q^{-\s^3D}
(1+\z\ab^*\s^++\z\ab\s^--\z^2q^{2D}\tau^+).
\en
Apart from the power $\z^{-1/2}$,
the $L$ operator is a quadratic polynomial in $\z$, and its inverse has
a pole at $\z^2=1$.

The total spin operator which commutes with the $L$ operator is
\be
S^\pm=\mp D_A+S,\ [S^\pm,L_{A,j}^\pm(\z)]=0.
\en
This suggests a construction of an operator similar to
$\tb(\z,\al)$ using the trace on $W^\pm$, in which $q^{-\al\s^3}$ is replaced by
$q^{\pm2\al D}$. This construction leads to the $Q$ operators acting on
$\mathcal V$:
\bea
&&\bQ^\pm(\z,\al)(X)=\pm(1-q^{\pm2(\al-\bS)})\z^{\pm(\al-\bS)}
\Tr^\pm_A\left(q^{\pm2\al D_A}T^\pm_A(\z)^{-1}XT^\pm_A(\z)\right),\label{Q}\\
&&T_A^\pm(\z)=L^\pm_{A,n}(\z/\xi_n)\cdots L^\pm_{A,1}(\z/\xi_1).
\nn
\ena
Here the trace functional
\be
\Tr^\pm_A(q^{\pm2\al D_A}\ \cdot):Osc_A\rightarrow\C(q^{2\al})
\en
is defined to be zero on each spin sector
$Osc^{(s)}=\{x\in Osc\,|\,[D,x]=sx\}$ with $s\neq 0$,  
and on the spin zero sector $Osc^{(0)}$ 
it is defined by
\be
\Tr^\pm_A(q^{\pm2\al D_A}q^{mD_A})=\pm(1-q^{\pm2\al+m})^{-1}.
\en
Since
$T^\pm_A(\z)^{-1}XT^\pm_A(\z)\bigl|_{\z=0}=q^{\mp2\bS D_A}(X)$, we have
$\z^{\mp(\al-\bS)}\bQ^\pm(\z,\al)\bigl|_{\z=0}={\rm id}_{\mathcal V}$.
This explains the prefactor $\pm(1-q^{\pm2(\al-\bS)})$
in the definition of $\bQ^\pm$.

The trace functional satisfies
\be
\Tr^+(q^{2\al D}x)=-\Tr^-(q^{2\al D}x).
\en
{}From this follows
\be
\jb\cdot\bQ^+(\z,\al)=\bQ^-(\z,\al)\cdot\jb.
\en
The Yang-Baxter relation for the $L$ operator 
\be
R_{i,i+1}(\z_i/\z_{i+1})L_{A,i}(\z_i)L_{A,i+1}(\z_{i+1})
=L_{A,i+1}(\z_{i+1})L_{A,i}(\z_i)R_{i,i+1}(\z_i/\z_{i+1})
\en
implies the equivariance.
\be
\sb_i\cdot\bQ^\pm(\z,\al)=\bQ^\pm(\z,\al)\cdot\sb_i.
\en
\subsection{Baxter's $TQ$ relation}
The following proposition is due to \cite{BLZ}. 
We give here a proof, for the purpose of introducing 
further notation and formulas.
\begin{prop}
\bea
\tb(\z,\al)\bQ^\pm(\z,\al)=\bQ^\pm(q^{-1}\z,\al)+\bQ^\pm(q\z,\al).
\label{TQ}
\ena
\end{prop}
\begin{proof}
Define the triangular transfer matrix
\bea
&&T^+_{\{A,a\}}(\z)=(G_{A,a}^+)^{-1}T^+_A(\z)T_a(\z)G_{A,a}^+,
\label{fusion}
\\
&&G_{A,a}^+=q^{-\s^3_aD_A}(1+\ab^*_A\s^+_a).
\ena
The triangularity is local:
\bea
&&L^+_{\{A,a\},j}(\z)=(G_{A,a}^+)^{-1}L^+_{A,j}(\z)R_{a,j}(\z)G_{A,a}^+
\label{TRIG}
\\
&&\quad=
(q\z-q^{-1}\z^{-1})\cdot
\begin{pmatrix}
L^+_{A,j}(q^{-1}\z)q^{-\s^3_j/2}&0\\
\g(\z)L^+_{A,j}(q\z)\s^+_jq^{-2D_A+1/2}&\beta(\z)L^+_{A,j}(q\z)q^{\s^3_j/2}
\end{pmatrix}_a\,.
\nn
\ena
The inverse is given by
\be
&&L^+_{\{A,a\},j}(\z)^{-1}=
\frac{1}{q\z-q^{-1}\z^{-1}}
\\
&&\quad\times
\begin{pmatrix}
q^{\s^3_j/2}L^+_{A,j}(q^{-1}\z)^{-1}&0\\
-\g(q^{-1}\z)\s^+_jq^{-2D_A-1/2}L^+_{A,j}(q^{-1}\z)^{-1}
&\beta(\z)^{-1}q^{-\s^3_j/2}
L^+_{A,j}(q\z)^{-1}
\end{pmatrix}_a\,.
\en
Using the commutativity
\be
[q^{2\al D_A-\al\s^3_a},G_{A,a}^+]=0,
\en
we obtain the $TQ$ relation.
The effect of the shift $\z^{\al-\bS}\rightarrow (q^{\mp1}\z)^{\al-\bS}$
is cancelled by $q^{\mp\s^3_j/2}$ in the diagonal elements of 
$L^+_{\{A,a\},j}(\z)^{\pm1}$ and $q^{-\al\s^3_a}$ in $\tb(\z,\al)$.
\end{proof}
In the $TQ$ relation,  
the order of the product $\tb(\z,\alpha)\bQ^\pm(\z,\al)$
can be reversed.
We can show
\be
\bQ^\pm(\z,\al)\tb(\z,\al)=\bQ^\pm(q\z,\al)+\bQ^\pm(q^{-1}\z,\al).
\en
The argument is quite parallel.
Set\footnote{Notice the signs $\pm$ in $T^{*+}_{\{a,A\}}$ and $T^-_A$;
they are {\it not} misprints.}
\be
&& T^{*+}_{\{a,A\}}(\z)=(F_{A,a}^+)^{-1}T_a(\z)T^-_A(\z)F_{A,a}^+,\\
&&F_{A,a}^+=1-\ab_A\s^+_a,\
[q^{-2\al D_A-\al\s^3_a},F_{A,a}^+]=0.
\en
We have
\be
L^{*+}_{\{a,A\},j}(\z)
&\buildrel{\rm def}\over=&(F_{A,a}^+)^{-1}R_{a,j}(\z)L^-_{A,j}(\z)F_{A,a}^+
\\
&=&
(q\z-q^{-1}\z^{-1})\cdot
\begin{pmatrix}
\beta(\z)L^-_{A,j}(q\z)q^{-\s^3_j/2}&0\\
\g(\z)\s^+_jL^-_{A,j}(q\z)q^{-\s^3_j/2}&L^-_{A,j}(q^{-1}\z)q^{\s^3_j/2}
\end{pmatrix}_a
\en
and
\be
&& L^{*+}_{\{a,A\},j}(\z)^{-1}
\\
&&\quad
=
\frac{1}{q\z-q^{-1}\z^{-1}}\cdot
\begin{pmatrix}
\beta(\z)^{-1}q^{\s^3_j/2}L^-_{A,j}(q\z)^{-1}&0\\
-\g(q^{-1}\z)q^{-\s^3_j/2}L^-_{A,j}(q^{-1}\z)^{-1}\s^+_j&
q^{-\s^3_j/2}L^-_{A,j}(q^{-1}\z)^{-1}
\end{pmatrix}_a.
\en
Using these quantities we arrive at 
\be
\bQ^-(\z,\al)\tb(\z,\al)=\bQ^-(q\z,\al)+\bQ^-(q^{-1}\z,\al).
\en
Later on we will also use 
\be
&&
L^-_{\{A,a\},j}(\z)=\sigma^1_a\sigma^1_jL^+_{\{A,a\},j}(\z)\sigma^1_a\sigma^1_j\,,
\\
&&
L^{*-}_{\{A,a\},j}(\z)=\sigma^1_a\sigma^1_jL^{*+}_{\{A,a\},j}(\z)\sigma^1_a\sigma^1_j\,.
\en

\subsection{Conjugate transfer matrices}
We define conjugate transfer matrices by reversing the operations
inside the trace.
\be
&&\tb^*(\z,\al)(X)=\tr_a(T_a(\z)q^{\al\s^3_a}XT_a(\z)^{-1})\,,
\\
&&\bQ^{*\pm}(\z,\al)(X)=\pm(1-q^{\pm2(\al-\bS)})\z^{\pm(\al-\bS)}
\Tr^\pm_A\left(T^\mp_A(\z)q^{\pm2\al D_A}XT^\mp_A(\z)^{-1}\right)\,.
\en
They satisfy the same $TQ$ relations
\bea
\tb^*(\z,\al)\bQ^{*\pm}(\z,\al)=\bQ^{*\pm}(q^{-1}\z,\al)+\bQ^{*\pm}(q\z,\al)
\label{TQ*}
\ena
and the equivariance with respect to $\sb_i$ and $\jb$.

\subsection{Reduction property}
Now, we consider the extension of the interval from $[1,n]$
to $[1,n+1]$ or $[0,n]$. 
We exhibit the dependence on the interval of various quantities
by a suffix, and write $\cV_{[1,n]}$, $\tb_{[1,n]}(\z,\al)$ and so on. 
The operator $\tb(\z,\al)$ has the reduction property to the right:
\be
\tb_{[1,n+1]}(\z,\al)(X_{[1,n]}\otimes1)
=\tb_{[1,n]}(\z,\al)(X_{[1,n]})\otimes1.
\en
This is tautological because of the inverse adjoint action:
in the left hand side, the adjoint action 
$R_{a,n+1}^{-1}(\z,\al)XR_{a,n+1}(\z,\al)$
on the quantum space $n+1$ appears at the innermost place. 
On the other hand,
the operator $\tb^*(\z,\al)$ has the reduction property to the left:
\be
\tb^*_{[0,n]}(\z,\al)(q^{\al\s^3}\otimes X_{[1,n]})
=q^{\al\s^3}\otimes\tb^*_{[1,n]}(\z,\al)(X_{[1,n]}).
\en
This is also tautological because 
the operation appearing at the innermost place 
is $R_{a,0}(\z,\al)XR_{a,0}(\z,\al)^{-1}$, 
which commutes with the multiplication by $q^{\al(\s^3_a+\s^3_0)}$.

In \cite{BJMST} were constructed
operators which satisfy reduction to both ends.
In order to realize such operators, we have to modify the form of the 
reduction.
We do not change the reduction to the right. We change the reduction
to the left in the form
\be
\cb^\pm_{[0,n]}(q^{\al\s^3}\otimes X_{[1,n]})
=q^{(\al\mp1)\s^3}\otimes\cb^\pm_{[1,n]}(X_{[1,n]}).
\en
{}For this purpose
we consider the off-diagonal blocks in the monodromy matrices.
\subsection{Operators in off-diagonal blocks}
We define a $q$ difference operator with respect to the spectral variable
\be
\Delta_q(F(\z))=F(q\z)-F(q^{-1}\z).
\en
A function of this form is said to be {\it $q$-exact}. 

\begin{prop}
Set
\bea
&&\bk^\pm(\z,\al)(X)
\label{bk}
\\
&&=\pm(1-q^{\pm2(\al-\bS)})\z^{\pm(\al-\bS)}
\Tr^\pm_A\tr_a\left(q^{\pm2(\al\mp1)D_A}\s^\pm_a
T_a(\z)^{-1}T^\pm_A(\z)^{-1}XT^\pm_A(\z)T_a(\z)\right)
\nn\\
&&=\pm(1-q^{\pm2(\al-\bS)})\z^{\pm(\al-\bS)}
\Tr^\pm_A\tr_a\left(q^{\pm2\al D_A}\s^\pm_a
T^\pm_{\{A,a\}}(\z)^{-1}XT^\pm_{\{A,a\}}(\z)\right)\,.
\nn
\ena
This operator satisfies the reduction property modulo a $q$-exact form:
\bea
&&\bk^\pm_{[1,n]}(\z,\al)(q^{\al\s^3_1}X_{[2,n]})
=q^{(\al\mp1)\s^3_1}\bk^\pm_{[2,n]}(\z,\al)(X_{[2,n]})
\label{REDUCTION}\\
&&+\s^\pm_1\Delta_q\left(\frac{q-q^{-1}}{\z/\xi_1-(\z/\xi_1)^{-1}}
\bQ^\pm_{[2,n]}(\z,\al\mp1)(X_{[2,n]})\right)\,.
\nn
\ena
\end{prop}
\begin{proof}
We subdivide the quantum space as $\cV=\cV_{1}\otimes \cV_{[2,n]}$,  
where $\cV_{1}=\End(V_1)$ and $\cV_{[2,n]}=\End(V_2\otimes\cdots\otimes V_n)$.
Let us calculate
\be
C&=&(1-q^{2(\al-\bS)})^{-1}\z^{-(\al-\bS)}
\bk^+(\z,\al)(q^{\al\s^3_1}X_{[2,n]})\\
&=&\Tr^+_A\tr_a\left(q^{2(\al-1)D_A}\s^+_a
T_a(\z)^{-1}T^+_A(\z)^{-1}q^{\al\s^3_1}X_{[2,n]}T^+_A(\z)T_a(\z)\right)
\en
where $X_{[2,n]}\in\cV_{[2,n]}$. Using the commutativity of the $L$ operator
and the total spin operator, we have
\be
&&C=q^{(\al-1)(\s^3_1-1)}\Tr^+_A\tr_a\bigl(\s^+_a
R_{a,1}(\z/\xi_1)^{-1}L^+_{A,1}(\z/\xi_1)^{-1}
q^{\s^3_1}q^{(\al-1)(2D_A-\s^3_a)}\\
&&\times T_{a,[2,n]}(\z)^{-1}T^+_{A,[2,n]}(\z)^{-1}
X_{[2,n]}T^+_{A,[2,n]}(\z)T_{a,[2,n]}(\z)L^+_{A,1}(\z/\xi_1)R_{a,1}(\z/\xi_1)\bigr)
\en
Note that $\left(G^+_{A,a}\right)^{-1}\s^+_aG^+_{A,a}=\s^+_aq^{2D_A}$.
Using \eqref{TRIG} we rewrite as
\be
&&C=q^{(\al-1)(\s^3_1-1)}\Tr^+_A\tr_a\bigl(\s^+_aq^{2D_A}
L^+_{\{A,a\},1}(\z/\xi_1)^{-1}q^{\s^3_1}q^{(\al-1)(2D_A-\s^3_a)}\\
&&\times T^+_{\{A,a\},[2,n]}(\z)^{-1}
X_{[2,n]}T^+_{\{A,a\},[2,n]}(\z)L^+_{\{A,a\},1}(\z/\xi_1)\bigr)\,.
\en
Collecting the terms containing operators acting on $\cV_1$, we obtain
\be
&&C=q^{(\al-1)(\s^3_1-1)}\Tr^+_A\tr_a
\bigl(
\bigl\{L^+_{\{A,a\},1}(\z/\xi_1)^{t_1}\s^+_aq^{2D_A}
(L^+_{\{A,a\},1}(\z/\xi_1)^{-1}q^{\s^3_1})^{t_1}\bigr\}^{t_1}\\
&&\times q^{(\al-1)(2D_A-\s^3_a)}
T^+_{\{A,a\},[2,n]}(\z)^{-1}X_{[2,n]}T^+_{\{A,a\},[2,n]}(\z)\bigr)\,.
\en
The part $\{\cdots\}^{t_1}$ reads as
\be
\{\cdots\}^{t_1}=q^{2D_A}\s^+_a
+\frac{q-q^{-1}}{q\z/\xi_1-(q\z/\xi_1)^{-1}}\s^+_1\tau^-_a
-\frac{q-q^{-1}}{q^{-1}\z/\xi_1-q(\z/\xi_1)^{-1}}\s^+_1\tau^+_a.
\en
Using \eqref{Q} we obtain \eqref{REDUCTION}.
The term $q^{2D_A}\s^+_a$ creates the main term and the rest creates
the $q$-exact term.
The presence of the factor $(1-q^{\al-\bS})$ in 
$\bk^+$ and $\bQ^+$
does not effect the form of the equality because
$\bS_{[1,n]}(\s^+_1X_{[2,n]})=\s^+_1(\bS_{[2,n]}+1)(X_{[2,n]})$.
On the other hand,
the factor $\z^{\al-\bS}$ matters because it does not commute with the
$q$ difference operator. It adjusts the effect of 
$\tau^\pm_a\s^3_a=\pm\tau^\pm_a$ 
in
the factor $q^{(\al-1)(2D_A-\s^3_a)}$.
\end{proof}
The operators $\bk^\pm(\z,\al)$ enjoy
the non-trivial reduction property to the left
with a $q$-exact form as remainder term. 
In the next subsection 
we define the operators $\cb^\pm_k(\z,\al)$ by taking
the singular part at $\z=\xi_k$. 
The operator $\bQ^\pm_{[2,n]}(\z,\al\mp1)$
has no pole at $\z=q^{\pm1}\xi_k$. 
Therefore, the $q$-exact term has no pole at $\z=\xi_k$. It implies that 
the operators $\cb^\pm_k(\z,\al)$ satisfy the reduction property
without remainder terms.

Next we define operators with non-trivial reduction to the right.
We demand the tautological reduction to the left in the form
\be
\bk^{*\pm}_{[1,n]}(q^{(\al\pm1)\s^3_1}X_{[2,n]})=
q^{\al\s^3_1}\bk^{*\pm}_{[2,n]}(X_{[2,n]}).
\en
This is achieved by defining 
\be
&&\bk^{*\pm}(\z,\al)(X)
=\pm(1-q^{\pm2(\al\pm1-\bS)})\z^{\pm(\al\pm1-\bS)}\\
&&\times\Tr^\pm_A\tr_a\left(\s^\pm_aT_a(\z)T^\mp_A(\z)
q^{\al(\pm2D_A+\s^3_a)\mp2S}\,
X\,T^\mp_A(\z)^{-1}T_a(\z)^{-1}\right)\,.
\en
In this definition, we make the shift $q^{\mp2S}X$ because
we want to have the reduction to the left as above.
As for the reduction to the right we have
\begin{prop}
\be
&&\bk^{*\pm}_{[1,n]}(\z,\al)(X_{[1,n-1]})
=\bk^{*\pm}_{[1,n-1]}(\z,\al)(X_{[1,n-1]})\\
&&+\s^\pm_n
\Delta_q\left(\frac{q-q^{-1}}{\z/\xi_n-(\z/\xi_n)^{-1}}
\bQ^{*\pm}_{[1,n-1]}(\z,\al)(q^{\mp2S_{[1,n-1]}}X_{[1,n-1]})
\right).
\en
\end{prop}
\begin{proof}
Let us calculate
\be
&&
\Tr^+_A\tr_a\left(\s^+_aT_a(\z)T^-_A(\z)
q^{\al(2D_A+\s^3_a)-2S_{[1,n]}}X_{[1,n-1]}T^-_A(\z)^{-1}T_a(\z)^{-1}\right)\\
&&\quad=\Tr^+_A\tr_a\left(\s^+_aT^{*+}_{\{a,A\}}(\z)
q^{\al(2D_A+\s^3_a)-2S_{[1,n]}}X_{[1,n-1]}T^{*+}_{\{a,A\}}(\z)^{-1}\right)\\
&&\quad=\Tr^+_A\tr_a\Bigl(\left\{(q^{-\s^3_n}
L^{*+}_{\{a,A\},n}(\z/\xi_n)^{-1})^{t_n}\s^+_a
L^{*+}_{\{a,A\},n}(\z/\xi_n)^{t_n}\right\}^{t_n}\\
&&\quad\times T^{*+}_{\{a,A\},[1,n-1]}(\z)
q^{\al(2D_A+\s^3_a)-2S_{[1,n-1]}}X_{[1,n-1]}
T^{*+}_{\{a,A\},[1,n-1]}(\z)^{-1}\Bigr)\,.
\en
The statement of the proposition follows from
\be
&&\left\{(q^{-\s^3_n}L^{*+}_{\{a,A\},n}(\z)^{-1})^{t_n}\s^+_a
L^{*+}_{\{a,A\},n}(\z)^{t_n}\right\}^{t_n}\\
&&=\s^+_a+\frac{q-q^{-1}}{q\z-(q\z)^{-1}}\s^+_n\tau^+_a
-\frac{q-q^{-1}}{q^{-1}\z-q\z^{-1}}\s^+_n\tau^-_a.
\en
\vskip-18pt
\end{proof}
\subsection{Operators $\bb$ and $\cb$}
The operator $\tb(\z,\al)$ does not have a pole at $\z^2=\xi_j^2$, but
has a pole at $\z^2=q^{\pm2}\xi_j^2$. 
The operators $\bQ^\pm(\z,\al)$ do not have a pole at $\z^2=q^{\pm2}\xi_j^2$,  
but have a pole at $\z^2=\xi_j^2$. 
These are consequences of the pole structures
of $R_{a,j}(\z)^{\pm1}$ and $L_{A,j}(\z)^{\pm1}$.
Consider the matrix
$T^\pm_{\{A,a\}}(\z)^{-1}XT^\pm_{\{A,a\}}(\z)$, which is triangular
in the auxiliary space $V_a$. The diagonal part does not have a pole at
$\z^2=\xi_j^2$ because poles cancel in the adjoint action.
This cancellation breaks down for the off-diagonal elements.
The operators $\cb^\pm(\z,\al)$ do have a simple pole at $\z^2=\xi_j^2$.

Set
\be
&&\cb^-(\z,\al)=\bb(\z,\al)=(1-q^{2(\al-\bS)})^{-1}{\rm sing}\,
\bk^-(\z,\al),\\
&&\cb^+(\z,\al)=\cb(\z,\al)=q^{\al-\bS}{\rm sing}\,
\bk^+(\z,\al).
\en
The symbol ${\rm sing}$ means taking the singular part at
$\z=\xi_i$ $(i=1,\ldots,n)$. Therefore, we have a partial fraction
\be
\cb^\pm(\z,\al)=-\frac12\sum_{k=1}^n\frac{\xi_k}{\z-\xi_k}\cb^\pm_k(\al).
\en
The $q$ dependent normalization factors in the definition
do not alter the reduction properties. Taking the singular part in 
\eqref{REDUCTION} we obtain 

\begin{prop}\label{prop:c-red}
We have
\be
&&\cb^\pm_{[1,n]}(\z,\al)(X_{[1,n-1]})
=\cb^\pm_{[1,n-1]}(\z,\al)(X_{[1,n-1]}),\\
&&\cb^\pm_{[1,n]}(\z,\al)(q^{\al\s^3_1}X_{[2,n]})
=q^{(\al\mp1)\s^3_1}\cb^\pm_{[2,n]}(\z,\al)(X_{[2,n]}).
\en
\end{prop}

The normalization factors are so chosen that we have the Grassmann relation
\footnote{One can choose the normalization so that the equivariance
$\cb^+\leftrightarrow\cb^-$ with respect
to the spin reversal is valid. 
We do not exploit this possibility
because we will not use this property, and also because
it would require introduction of a square root.}.

\begin{prop}
\be
\cb^{\varepsilon_1}(\z_1,\al-\varepsilon_2)\cb^{\varepsilon_2}(\z_2,\al)=
-\cb^{\varepsilon_2}(\z_2,\al-\varepsilon_1)\cb^{\varepsilon_1}(\z_1,\al).
\en
\end{prop}
The proof is given in Appendix \ref{sec:app2}.

It follows from Proposition \ref{prop:c-red} that the residue
$\cb^\pm_n(\al)$ of $\cb^\pm_{[1,n]}(\z,\al)$ taken 
at the right end of the interval $\z=\xi_n$ 
annihilate states of the form $X_{[1,n-1]}$.
This operator is given explicitly as follows. 

\begin{prop}\label{PROP1}
Let 
$X\in\cV_{[1,n]}$. 
Then we have
\be
&&\cb^+_n(\al)(X)=
\frac{(1-q^{2(\al-\mathbb S)})(q\xi_n)^{\alpha-\mathbb S}}
{\prod_{j=1}^{n-1}\beta(\xi_n/\xi_j)}
\tilde{\bQ}^+(\xi_n,\al)
\left(({\bf a},1)_nX\begin{pmatrix}1\\-{\bf a}\end{pmatrix}_n\right),\\
&&\cb^-_n(\al)(X)=
\frac{(q^2\xi_n)^{-(\alpha-\mathbb S)}}
{\prod_{j=1}^{n-1}\beta(\xi_n/\xi_j)}
\tilde{\bQ}^-(\xi_n,\al)
\left((1,{\bf a})_nX\begin{pmatrix}-{\bf a}\\1\end{pmatrix}_n\right)
\en
where for 
$\tilde X\in Osc\otimes\cV_{[1,n-1]}$ we set
\be
&&
\tilde{\bQ}^\pm(\z,\al)(\tilde X)
=
q^{\mp S_{[1,n-1]}}
{\rm Tr}^\pm_A\left(q^{\pm2\al D_A}
T^\pm_{A,[1,n-1]}(q\z)^{-1}\tilde X\,T^\pm_{A,[1,n-1]}(q^{-1}\z)
\right)
q^{\mp S_{[1,n-1]}}\,.
\en
In particular, we have $\cb^\pm_n(\al)(X)\in \cV_{[1,n-1]}$.
\end{prop}
\begin{proof}
Consider $\bk^\pm(\z,\al)$ given by \eqref{bk}.
Because of the triangular structure of $L^\pm_{\{A,a\},j}(\z)$,
${\rm tr}_a\left(\s^\pm_aT^\pm_{\{A,a\}}(\z)^{-1}(X)T^\pm_{\{A,a\}}(\z)\right)$
consists of $2n$ terms.
The singular part of $\bk^\pm(\z,\al)$ at $\z=\xi_n$ comes only from the factor
$\gamma(q^{-1}\z/\xi_n)$ and $\beta(\z/\xi_n)^{-1}$ in
$L^\pm_{\{A,a\},n}(\z/\xi_n)^{-1}$.
Note also that $\beta(\z/\xi_n)$ is zero at $\z=\xi_n$. {}From these properties,
it follows that only two terms have singularities at $\z=\xi_n$.
Summing them up, we obtain the above expressions.
\end{proof}

To simplify the notation, let us use the following convention. 
{}For each $\beta\in\al+\Z$ 
take a copy 
$\mathcal V_\beta\simeq\mathcal V$, 
and define an operator $\alb$ acting on
$\mathcal V_{[\al]}=\oplus_{\beta\in\al+\Z}\mathcal V_\beta$ by 
\be
\alb|_{\mathcal V^{(s)}_\beta}
=\beta\,{\rm id}_{\mathcal V^{(s)}_\beta}.
\en
Let $I\subset\Z$ be a finite interval. We consider the quantum space
$\mathcal V_{[\al],I}$ by changing $[1,n]$ to $I$.
If $J=[j_1,j_2]$ ($i_1\leq j_1\leq j_2\leq i_2$)
is a subinterval of $I=[i_1,i_2]$, 
then $\mathcal V_{[\al],J}$ is naturally embedded in 
$\mathcal V_{[\al],I}$:
\be
\iota_{J,I}:\mathcal V_{\beta,J}\rightarrow\mathcal V_{\beta,I},\
X_{[j_1,j_2]}\mapsto
q^{\sum_{j=i_1}^{j_1-1}\beta\s^3_j}
X_{[j_1,j_2]}.
\en
We often drop $\iota_{J,I}$ when the meaning is clear without
writing it explicitly.

In this notation, the basic properties of the operators 
$\cb^+(\z)=\cb(\z)$ 
and $\cb^-(\z)=\bb(\z)$ are summarized as follows. 
\bea
&&\cb^{\varepsilon_1}(\z_1)\cb^{\varepsilon_2}(\z_2)=
-\cb^{\varepsilon_2}(\z_2)\cb^{\varepsilon_1}(\z_1),\label{GRASS}\\
&&\sb_i\,\cb^\pm(\z)=\cb^\pm(\z)\sb_i,\label{INV}\\
&&\cb^\pm_{[1,n]}(\z)(X_{[1,n-1]})
=\cb^\pm_{[1,n-1]}(\z)(X_{[1,n-1]}),\label{LREDUC}\\
&&\cb^\pm_{[1,n]}(\z)(q^{\scalb\s^3_1}X_{[2,n]})
=q^{\scalb\s^3_1}\cb^\pm_{[2,n]}(\z)(X_{[2,n]}).\label{RREDUC}
\ena

\section{Free Fermion Point}\label{sec:3}

In this section, 
we discuss the free fermion case $q=i$. 
The main point is that 
the operators $\bb_k$ and $\cb_k$ are cubic in appropriate 
free fermions (see Theorem \ref{thm:c} below). 

\subsection{Jordan-Wigner transformation in adjoint action} 
We continue to work with $\mathcal V=\End(V_1\otimes\cdots\otimes V_n)$.
Define the Jordan Wigner transformation
\be
\psi^\pm_j=\s^\pm_ji^{\mp\sum_{l=1}^{j-1}\s^3_l}\in\mathcal V.
\en
We define fermions\footnote{For $X\in\mathcal V^{(s)}$,
the operator $\Phi^\pm_j(\al)$ in this paper corresponds to
$\Phi^\pm_{\al-s+1,j}$ in \cite{BJMST}.} acting on $X\in\mathcal V^{(s)}$.
\be
&&\Psi^\pm_j(X)=\psi^\pm_jX-(-1)^sX\psi^\pm_j,\\
&&\Phi^\pm_j(\al)(X)=\frac1{1+i^{\mp2(\al-s)}}
(\psi^\pm_jX+i^{\mp2\al}X\psi^\pm_j).
\en
Define further the operator $\Psi^\pm_j$ (resp. $\Phi^\pm_j$) 
on $\mathcal V_{[\al]}$ 
whose restriction to $V^{(s)}_\beta$ (resp. $V^{(s)}_{\beta\pm1}$) is given by 
\be
\Psi^\pm_j:
\mathcal V^{(s)}_\beta\rightarrow \mathcal V^{(s\pm1)}_{\beta\mp1},
\quad \Phi^\pm_j(\beta):
\mathcal V^{(s)}_{\beta\pm1}\rightarrow \mathcal V^{(s\pm1)}_{\beta}.  
\en
They satisfy the canonical anti-commutation relations
\be
[\Psi^\epsilon_j,\Psi^{\epsilon'}_{j'}]_+=0,\
[\Phi^\epsilon_j,\Phi^{\epsilon'}_{j'}]_+=0,\
[\Psi^\epsilon_j,\Phi^{\epsilon'}_{j'}]_+=
\delta_{\epsilon+\epsilon',0}\delta_{j,j'}.
\en
When $q=i$, there is certain redundancy in considering
$\mathcal V_{\beta}$ for all $\beta\in\Z+\alpha$, 
since we have the periodicity 
$\Phi^\pm_k(\al+2)=\Phi^\pm_k(\al)$, 
$\tb(\z,\al+4)=\tb(\z,\al)$. 
Nevertheless, we choose to retain this convention
in order to keep contact with the generic case.   
Also $\bQ^\pm(\z,\al)$ is not quite periodic due to the 
overall power $\z^{\pm(\al-\bS)}$.

\begin{prop}\label{RED1}
The fermions $\Psi^\pm_{j,[1,n]}$ and $\Phi^\pm_{j,[1,n]}$
satisfy the reduction property to the both ends except for $j=1,n$.
\bea
&&\Psi^\pm_{j,[1,n]}(X_{[1,n-1]})
=(1-\delta_{j,n})\Psi^\pm_{j,[1,n-1]}(X_{[1,n-1]}),\label{R1}\\
&&\Psi^\pm_{j,[1,n]}(i^{\scalb\s^3_1}X_{[2,n]})
=i^{\scalb\s^3_1}\Psi^\pm_{j,[2,n]}(X_{[2,n]})\ (j\not=1),\label{R2}\\
&&\Phi^\pm_{j,[1,n]}(X_{[1,n-1]})
=
\Phi^\pm_{j,[1,n-1]}(X_{[1,n-1]})\ (j\not=n),\label{R3}
\\
&&\Phi^\pm_{j,[1,n]}(i^{\scalb\s^3_1}X_{[2,n]})
=(1-\delta_{j,1})i^{\scalb\s^3_1}\Phi^\pm_{j,[2,n]}(X_{[2,n]}).\label{R4}
\ena
\end{prop}
\begin{proof}
The reduction to the right \eqref{R1}, \eqref{R3} except for $j=n$ is
tautological. The case $j=n$ for \eqref{R1} goes
\be
\psi^+_nX_{[1,n-1]}-(-1)^sX_{[1,n-1]}\psi^+_n
=i^{-2S_{[1,n-1]}}\s^+_nX_{[1,n-1]}
-(-1)^sX_{[1,n-1]}i^{-2S_{[1,n-1]}}\s^+_n\,.
\en
Since $(-1)^{S_{[1,n-1]}}X_{[1,n-1]}=(-1)^sX_{[1,n-1]}(-1)^{S_{[1,n-1]}}$,
this is zero.
The reduction to the left \eqref{R2},\eqref{R4} except for $j=1$
follows from $\psi^\pm_{j,[1,n]}i^{(\al\pm1)\s^3_1}=
\psi^\pm_{j,[2,n]}i^{\al\s^3_1}$. The case $j=1$ goes
\be
\s^\pm_1i^{(\al\pm1)\s^3_1}X_{[2,n]}+i^{\mp2\al}
i^{(\al\pm1)\s^3_1}X_{[2,n]}\s^\pm_1
=\s^+_1i^{-1\mp\al}X_{[2,n]}+i^{\mp2\al}i^{1\pm\al}X_{[2,n]}\s^\pm_1
=0\,.
\en
\end{proof}
The free fermions $\Psi^\pm_j,\Phi^\pm_j$ do not satisfy the equivariance
with respect to the symmetric group action. In the next subsection,
we transform them to those which enjoy the equivariance.
\subsection{Equivariant fermions}
Set
\be
\Psih^\pm(\z)=\sum_{j=1}^n\Psi^\pm_j\frac{i^j\g(\z/\xi_j)}
{\prod_{l=1}^j\beta(\z/\xi_l)},\
\Phih^\pm(\z)=\sum_{j=1}^n\Phi^\pm_j\frac{i^j\g(\z/\xi_j)}
{\prod_{l=1}^j\beta(\z/\xi_l)}.
\en

\begin{prop}\label{INVFERMI}
The operators $\Psih^\pm(\z),\Phih^\pm(\z)$ commute 
with the action of the symmetric group $\mathfrak S_n$, 
\be
\sb_i\Psih^\pm(\z)=\Psih^\pm(\z)\sb_i,\
\sb_i\Phih^\pm(\z)=\Phih^\pm(\z)\sb_i.
\en
\end{prop}
\begin{proof}
The statement follows from the identity
\be
(\sb_1-1)\left(\s^+_1\frac{\g(\z/\xi_1)}{\beta(\z/\xi_1)}+
\s^+_2\s^3_1\frac{\g(\z/\xi_2)}{\beta(\z/\xi_1)\beta(\z/\xi_2)}\right)=0.
\en
\end{proof}
Set
\bea
\Psih^\pm_k=-{\rm res}_{\z=\xi_k}\Psih^\pm(\z)\frac{d\z}\z,\quad 
\Phih^\pm_k=-i\Phih^\pm(i\xi_k).\label{FERMIONS}
\ena
Proposition \ref{INVFERMI} implies the equivariance
\bea
\sb_i\Psih^\pm_k=\Psih^\pm_{s_i(k)}\sb_i,\quad
\sb_i\Phih^\pm_k=\Phih^\pm_{s_i(k)}\sb_i\,,\label{COV1}
\ena
where $s_i=(i,i+1)$ is the transposition. 
They are upper or lower triangular linear combinations of
$\Psi^\pm_j$ or $\Phi^\pm_j$.
\be
&&\Psih^\pm_k=\sum_{j=k}^n\Psi^\pm_j\frac{i^{j-1}\g(\xi_k/\xi_j)}
{\prod_{1\le l\le j\atop l(\neq k)}\beta(\xi_k/\xi_l)},
\\
&&\Phih^\pm_k=\sum_{j=1}^k\Phi^\pm_j{i^{-j+1}\g(\xi_k/\xi_j)}
\prod_{l=1}^{j-1}\beta(\xi_k/\xi_l).
\en
\begin{prop}\label{prop:ANT1}
The fermions $\Psih^\pm_k,\Phih^\pm_k$ satisfy the canonical anti-commutation
relations.
\be
[\Psih^\epsilon_j,\Psih^{\epsilon'}_{j'}]_+=0,\
[\Phih^\epsilon_j,\Phih^{\epsilon'}_{j'}]_+=0,\
[\Psih^\epsilon_j,\Phih^{\epsilon'}_{j'}]_+=
\delta_{\epsilon+\epsilon',0}\delta_{j,j'}.
\en
\end{prop}
\begin{proof}
Define an $n\times n$ matrix $P=(P_{jk})$ by 
\bea
P_{jk}=
\begin{cases}
\displaystyle{\frac{2\xi_j/\xi_k}{1-\xi_j^2/\xi_k^2}\prod_{l=j+1}^{k-1}
\frac{1+\xi_j^2/\xi_l^2}{1-\xi_j^2/\xi_l^2}}&\hbox{ if $j<k$};\\
1&\hbox{ if $j=k$};\\0&\hbox{ if $j>k$}.
\end{cases}
\label{eq:P}
\ena
The inverse is of a similar form, 
\be
(P^{-1})_{jk}=\begin{cases}
\displaystyle{\frac{2\xi_k/\xi_j}{1-\xi_k^2/\xi_j^2}\prod_{l=j+1}^{k-1}
\frac{1+\xi_k^2/\xi_l^2}{1-\xi_k^2/\xi_l^2}}&\hbox{ if $j<k$};\\
1&\hbox{ if $j=k$};\\0&\hbox{ if $j>k$}.
\end{cases}
\en
The statement follows from the equalities
\be
\Psih^\pm_k=\prod_{l=1}^{k-1}\frac{1+\xi_k^2/\xi_l^2}{1-\xi_k^2/\xi_l^2}
\sum_{j=1}^n\Psi_j^\pm P_{kj},\
\Phih^\pm_k=\prod_{l=1}^{k-1}\frac{1-\xi_k^2/\xi_l^2}{1+\xi_k^2/\xi_l^2}
\sum_{j=1}^n\Phi_j^\pm(P^{-1})_{jk}.
\en
\end{proof}
Let
$N^\pm_k=\Phih^\pm_k\Psih^\mp_k$
be the number operators corresponding to the fermion creation operators
$\Phih^\pm_k$ and the fermion annihilation operators $\Psih^\mp_k$.
They satisfy the following reduction properties.
\be
&&N^\pm_{j,[1,n]}(X_{[1,n-1]})
=(1-\delta_{j,n})N^\pm_{j,[1,n-1]}(X_{[1,n-1]}),\\
&&N^\pm_{j,[1,n]}(i^{\al\s^3_1}X_{[2,n]})
=\begin{cases}
i^{\al\s^3_1}X_{[2,n]}&\hbox{ if $j=1$;}\\
i^{\al\s^3_1}N^\pm_{j,[2,n]}(X_{[2,n]})&\hbox{ if $j\geq2$.}
\end{cases}
\en
\subsection{Cubic form in fermions}
We introduce another set of free fermions in which the operators
$\bb_k,\cb_k$ are written as cubic monomials.
Set
\be
&&U^\pm_k=\prod_{l\not=k}\frac{(1-\xi_k^2/\xi_l^2)^{N^\pm_l}}
{(1+\xi_k^2/\xi_l^2)^{N^\mp_l}},\\
&&\chi^+_k=\Psih^+_kU^+_k\xi_k^{\scalb-\bS-1}i^{\scalb},\\
&&\chi^-_k=\Psih^-_kU^-_k\xi_k^{-\scalb+\bS-1}
(-1)^{N^-_k}
\frac{i^{-\bS+1}}{1+i^{2(\scalb-\bS)}},\\
&&\chi^{*+}_k=i^{\bS-1}(1+i^{2(\scalb-\bS)})
(-1)^{N^-_k}
\xi^{\scalb-\bS+1}_k(U^-_k)^{-1}
\Phih^+_k,\\
&&\chi^{*-}_k=i^{-\scalb}\xi^{-\scalb+\bS+1}_k(U^+_k)^{-1}\Phih^-_k.
\en
\begin{prop}
The operators $\chi^\pm_k,\chi^{*\pm}_k$ satisfy the equivariance with respect
to the symmetric group action
\bea
\sb_i\chi^\pm_k=\chi^\pm_{s_i(k)}\sb_i,\
\sb_i\chi^{*\pm}_k=\chi^{*\pm}_{s_i(k)}\sb_i,\label{COV2}
\ena
the canonical anti-commutation relations
\bea
[\chi^\varepsilon_k,\chi^{\varepsilon'}_{k'}]_+=0,
[\chi^{*\varepsilon}_k,\chi^{*\varepsilon'}_{k'}]_+=0,
[\chi^\varepsilon_k,\chi^{*\varepsilon'}_{k'}]_+
=\delta_{\varepsilon+\varepsilon',0}\delta_{k,k'},\label{ANT2}
\ena
and the reduction relations
\bea
&&\chi_{k,[1,n+1]}(X_{[1,n]})=\chi_{k,[1,n]}(X_{[1,n]})\ (1\leq k\leq n)\,,
\\
&&\chi_{k,[0,n]}(q^{\scalb\s^3_0}X_{[1,n]})
=q^{\scalb\s^3_0}\chi_{k,[1,n]}(X_{[1,n]})\ (1\leq k\leq n)\,,
\label{RED2}
\ena
where $\chi=\chi^\pm,\chi^{*\pm}$.
\end{prop}
\begin{proof}
The equivariance \eqref{COV2} follows from \eqref{COV1}.
The anti-commutation relation \eqref{ANT2} follows from Proposition 
\ref{prop:ANT1}.
The reduction \eqref{RED2} follows from Proposition \ref{RED1}.
In particular, we use $\Psi^\pm_n(X_{[1,n-1]})=0$ and
$\Phi^\pm_1(i^{\scalb\s^3_1}X_{[2,n]})=0$.
\end{proof}
Set $N_k^\pm(\al)=N_k^\pm\bigl|_{\mathcal{V}_\al}$. 
The following formulas are proved in Appendix \ref{sec:app2}.

\begin{thm}\label{thm:tQ}
The transfer matrix and the $Q$ operators are diagonalized by the
fermion operators \eqref{FERMIONS}.
\bea
&&\tb(\z,\al)=
(i^{\al-\bS}+i^{-\al+\bS})
\prod_{k=1}^n
\left(\frac{1-\z^2/\xi_k^2}{1+\z^2/\xi_k^2}
\right)^{N^+_k(\al+1)+N^-_k(\al+1)}\,,
\label{ft}
\\
&&\z^{\mp(\al-\bS)}\bQ^\pm(\z,\al)=\prod_{k=1}^n
\frac{(1+\z^2/\xi_k^2)^{N^\pm_k(\al+1)}}
{(1-\z^2/\xi_k^2)^{N^\mp_k(\al+1)}}\,,
\label{fQ}
\\
&&\tb^*(\z,\al)(X)=i^{2S}
(i^{\al-\bS}+i^{-\al+\bS})\prod_{k=1}^n
\left(\frac{1-\z^2/\xi_k^2}{1+\z^2/\xi_k^2}
\right)^{2-N^+_k(\al+1)-N^-_k(\al+1)}(i^{-2S}X)\,,
\label{ft*}
\\
&&\z^{\mp(\al-\bS)}\bQ^{*\pm}(\z,\al)(X)=i^{2S}
\prod_{k=1}^n\frac{(1+\z^2/\xi_k^2)^{1-N^\mp_k(\al+1)}}
{(1-\z^2/\xi_k^2)^{1-N^\pm_k(\al+1)}}
(i^{-2S}X)\,.
\label{fQ*}
\ena
\end{thm}

\begin{thm}\label{thm:c}
The operator $\cb^\pm_k$ is a cubic monomial in the free fermions.
\be
\cb^\pm_k=\chi^\pm_k\chi^\mp_k\chi^{*\pm}_k.
\en
\end{thm}

Theorem \ref{thm:c} shows that, 
though $\cb^\pm_k$ themselves are not fermions, 
they act in a very simple manner on a basis created by fermions. 
{}For later reference, let us record the formula for this basis. 
Set $\chi^{*0}_k=1$, 
$\chi^{*\overline0}_k=\chi^{*+}_k\chi^{*-}_k$. 
Define a basis $\{\chi_p^{(\beta)}\}$ of the space $\mathcal V_{\beta}$
indexed by sequences $p=(p_1,\ldots,p_n)$ with $p_j\in\{+,-,0,\overline0\}$ by 
\be
\chi_p^{(\beta)}=\varepsilon(p)
\chi^{*p_n}_n\cdots\chi^{*p_1}_1({\rm id}_{\beta+\sum_{j=1}^ns(p_j)}), 
\en
where  ${\rm id}_\beta\in\mathcal V_\beta$ is the identity operator, 
$s(\pm)=\pm1$, $s(0)=s(\bar{0})=0$, 
\bea
\varepsilon(p)=
(-1)^{\sharp\{(i,j)\mid i<j,\ p_i=-,p_j=+\}},
\label{epsilon}
\ena
and $\sharp(Z)$ the cardinality of a set $Z$. 

\begin{cor}
The action of $\bb_k,\cb_k$ is given by
\be
&&\bb_k(\chi^{(\beta)}_{p})=\delta_{p_k,+}\varepsilon^+(p,k)
\chi^{(\beta+1)}_{p[k]}\,,
\\
&&\cb_k(\chi^{(\beta)}_{p})=\delta_{p_k,-}\varepsilon^-(p,k)
\chi^{(\beta-1)}_{p[k]}\,,
\en
where
\bea
&&\varepsilon^+(p,k)=(-1)^{\sharp(\{i|p_i=-\})+\sharp(\{i\mid i>k,\ p_i=+\})},
\label{DEF3}
\\
&&\varepsilon^-(p,k)=(-1)^{\sharp(\{i\mid i>k,\ p_i=-\})},
\label{DEF1}
\\
&&p[k]_j=\begin{cases}
0&\hbox{ if $j=k;$}\\p_j&\hbox{ if $j\not=k$.}
\end{cases}
\label{DEF2}
\ena
\end{cor}
\medskip

{\it Example.}\quad It is simple to calculate 
\be
\chi_1^{*\overline0}(\id_{\al})=\chi^{*+}_{1,[1,1]}\chi^{*-}_{1,[1,1]}(\id_{\al})=
\frac{i^\al+i^{-\al}}{i^\al-i^{-\al}}\cdot i^{\al\sigma^3_1}. 
\en
Using this and 
the reduction relations, we see that a 
successive application of 
$\chi^{*\overline0}_k$ to the identity operator 
produces the vacuum states, 
\be
\chi^{(\al)}_{\tiny{(\underbrace{\overline0,\ldots,\overline0}_k,
\underbrace{0,\ldots,0}_{n-k})}}
=\left(\frac{i^\al+i^{-\al}}{i^\al-i^{-\al}}\right)^k
i^{\al\sum_{j=1}^k\s^3_j}\,.
\en

\section{Basis for generic $q$}\label{sec:4}

In this section we introduce an inductive limit $\cW_{[\al]}$ of
the space $\cV_{[\al],I}$ when the interval $I=[k,l]$ becomes
infinite, $k\rightarrow-\infty,l\rightarrow\infty$.
The operators $\cb^\pm_j(\al):\cW_\al\rightarrow\cW_{\al\mp1}$
act on this space. Generalizing the result in the $q=i$ case,
we prove the existence of a basis for generic $q$,
on which the actions of $\cb^\pm_j(\al)$ are simple.
\subsection{Inductive limit}
We work with an infinite lattice
fully equipped with the spectral parameters.
Let $K$ be the field of rational functions in infinite variables
$\xi_j$ $(j\in\Z)$ with coefficients in $\C(y)$ $(y=q^\alpha)$.
Actually, the field $K$ is an inductive limit
of the field of rational functions $K_I$ $(I=[k,l])$
in the variables $\xi_k,\ldots,\xi_l$
when $k\rightarrow-\infty$ and $l\rightarrow\infty$.
Let $\mathfrak S_\infty$ be the infinite symmetric group generated by
the elements $s_i$, which is the transposition of $i$ and $i+1$.
On $K$ there is an action of $\mathfrak S_\infty$
such that $s_i=r_{i,i+1}$.
There is also an automorphism, $\tau(\xi_j)=\xi_{j+1}$,
which corresponds to the shift of the lattice.
{}Finally, there is an automorphism, $\kappa(y)=qy$, which corresponds
to the shift of the disorder parameter $\al$.

We define the vector space $\mathcal W_{[\alpha]}$ over $K$ as
the inductive limit of the vector spaces $\mathcal V_{[\al],I}$
where the inclusion maps are given by $\iota_{J,I}$ for intervals $J\subset I$.
We denote the subspace $K\otimes\mathcal V_{[\al],I}\subset\mathcal W_{[\al]}$
by $\left(\mathcal W_{[\al]}\right)_{I}$.
The total spin operator $\mathbb S$ is well-defined on $\cW_{[\al]}$.
We have the decomposition
\be
\cW_{[\al]}=\oplus_{s\in\Z}\cW^{(s)}_{[\al]},\
\cW^{(s)}_{[\al]}=\{X\in\cW_{[\al]}|\mathbb S(X)=sX\}.
\en
Note also that
$\mathcal W^{(s)}_{[\al]}=\oplus_{\beta\in\al+\Z}\mathcal W^{(s)}_\beta$.

Schematically, an infinite tensor product of the form
\be
\cdots\otimes y^{\s^3}\otimes y^{\s^3}\otimes
\s^+\otimes1\otimes \s^-\otimes y^{\s^3}\otimes1\otimes 1\otimes\cdots
\en
represents a vector in $\mathcal W_{\al}$.
To be more precise, denote by $\mathcal{P}$ the set of all maps 
$p:\Z\rightarrow\{+,-,0,\overline0\}$, $j\mapsto p_j$, 
such that 
\be
\sharp(\{j<0|p_j\not=\overline0\})<\infty,\quad
\sharp(\{j>0|p_j\not=0\})<\infty.
\en
An element of $\mathcal P$ will be called {\it label}. 
We use labels 
to represent states in $\mathcal W_{\al+k}$:
\be
v^{(\al+k)}_p=\otimes_{j\in\Z}v^{(\al+k)}_{p_j},
\en
where $v^{(\al+k)}_\pm=\s^\pm$, $v^{(\al+k)}_0=1$,
$v^{(\al+k)}_{\overline 0}=(q^ky)^{\s^3}$.
The set of states $v^{(\beta)}_p$ is a basis of $\mathcal W_\beta$,
but this is not the one we are looking for.

\subsection{Actions on $\mathcal W_{[\al]}$}
The automorphism $\kappa$ induces an isomorphism $\kapb$
of $\mathcal W_{[\al]}$
such that $\kapb:\mathcal W_\beta\rightarrow\mathcal W_{\beta+1}$.
The automorphism $\tau$ induces an isomorphism $\taub$ 
of $\mathcal W_{[\al]}$ such that
$\taub:\mathcal W_{\beta,[k,l]}\rightarrow\mathcal W_{\beta,[k+1,l+1]}$.
The action of the symmetric group $\mathfrak S_n$
on $\mathcal V$ induces an action of the
infinite symmetric group $\mathfrak S_\infty$ on $\cW_{[\al]}$.
We denote the action of $s_i\in\mathfrak S_\infty$ by $\sb_i$. We have
\be
\kapb\cdot\taub=\taub\cdot\kapb,\
\kapb\cdot\sb_i=\sb_i\cdot\kapb,\
\taub\cdot\sb_i=\sb_{i+1}\cdot\taub.
\en
Let $\mathcal A$ be the Grassmann algebra over $\C$
generated by $c^\pm_i$ $(i\in\Z)$.
Because of the reduction property \eqref{LREDUC}, \eqref{RREDUC},
the action of the operators
$\cb^\pm_j(\al):\mathcal V_\al\rightarrow\mathcal V_{\al\mp1}$
induces an action
$\cb^\pm_j:\mathcal W_\al\rightarrow\mathcal W_{\al\mp1}$.
We extend this action to $\mathcal W_{[\al]}$ by requiring
\be
\kapb\cdot\cb^\pm_i=\cb^\pm_i\cdot\kapb.
\en
Then, because of \eqref{GRASS}, it gives
an action of $\mathcal A$ on $\mathcal W_{[\al]}$.
Because of \eqref{INV} and by an obvious reason, we have the equivariance
\bea
&&\sb_i\cdot\cb^\pm_j=\cb^\pm_{s_i(j)}\cdot\sb_i,\label{Cov1}\\
&&\taub\cdot\cb^\pm_j=\cb^\pm_{j+1}\cdot\taub.\label{Cov2}
\ena
We observed that there is an action of the
algebra $\widetilde\cA$ over $\C$ generated by $\kapb^{\pm1}$,
$\taub^{\pm1}$, $\sb_i$ and $\cb^\pm_j$.
This action is $\C$-linear and satisfies the equivariance
\be
&&\kapb(f\cdot X)=\kappa(f)\cdot\kapb(X),\\
&&\taub(f\cdot X)=\tau(f)\cdot\taub(X),\\
&&\sb_i(f\cdot X)=r_{i,i+1}(f)\cdot\sb_i(X),\\
&&\cb^\pm_j(f\cdot X)=f\cdot\cb^\pm_j(X)
\en
for $f\in K$ and $X\in\mathcal W_{[\al]}$.
\subsection{Space of labels}
Let $\C[\Z\times\cP]$ be the vector space with the basis $\Z\times\cP$.
Our strategy is first to define a representation of the algebra
$\widetilde\cA$ on $\C[\Z\times\cP]$,
and then to show the existence of an intertwiner
\be
\mathcal X:
\C[\Z\times\cP]\rightarrow\cW_{[\al]}.
\en
The vectors $X^{(\al+k)}_p=\mathcal X(k,p)\in\cW_{\al+k}$ $(k\in\Z,p\in\cP)$
constitute the wanted $K$ basis of $\cW_{[\al]}$.

There is an action of $\mathfrak S_\infty$ on the vector space
$\C[\cP]$ such that
\be
s_i(p)=\begin{cases}
-p&\hbox{ if $(p_i,p_{i+1})=(+,+)$ or $(-,-)$;}\\
(\ldots,p_{i+1},p_i\ldots)&\hbox{ otherwise.}
\end{cases}
\en
This representation is isomorphic to a direct sum of
the representations induced from the representation
\be
{\rm id}_{(-\infty,k-1]}\otimes{\rm sgn}_{[k,l-1]}
\otimes{\rm sgn}_{[l,m-1]}\otimes{\rm id}_{[m,\infty)}
\en
of the parabolic subgroup
\be
{\mathfrak S}_{(-\infty,k-1]}\times{\mathfrak S}_{[k,l-1]}
\times{\mathfrak S}_{[l,m-1]}\times{\mathfrak S}_{[m,\infty)}
\en
for all possible choices of $k\leq l\leq m$. 
Here 
${\mathfrak S}_{[k,l-1]}$ denotes the symmetric group of degree $l-k$ 
acting on the interval $[k,l-1]$, 
and ${\rm sgn}_{[k,l-1]}$ is its sign representation. 
Set
\be
\left(p^{(k,l,m)}\right)_j=\begin{cases}
\overline0&\hbox{ if $j\leq k-1$;}\\
+&\hbox{ if $k\leq j\leq l-1$;}\\
-&\hbox{ if $l\leq j\leq m-1$;}\\
0&\hbox{ if $m\leq j$.}
\end{cases}
\en
We identify the element $p^{(k,l,m)}\in\cP$  with the cyclic vector
of the induced representation.
The above action on $\C[\cP]$ is lifted to $K[\cP]$
by requiring the equivariance 
$s_i(f\cdot p)=r_{i,i+1}(f)\cdot s_i(p)$. 
Similarly, we
have a natural action $\tau(p)_j=p_{j-1}$ of the shift operator.
Next we define an action of the Grassmann algebra $\mathcal A$ on $\C[\cP]$.
{}For $p\in\cP$ set
\bea
c^\pm_k(p)=\delta_{p_k,\mp}\varepsilon^{\mp}(p,k)p[k].
\label{ACT}
\ena
Here we used \eqref{DEF1}--\eqref{DEF3} for $p\in\cP$.
One can check the equivariance of this action:
\be
&&s_ic^\pm_j=c^\pm_{s_i(j)}s_i,
\\
&&\tau c^\pm_j=c^\pm_{j+1}\tau.
\en
{}Finally, we define the action of $\widetilde\cA$ on $\C[\Z\times\cP]$.
The action of $\kappa$ is such that
$\kappa(k,p)=(k+1,p)$. The actions of the other operators are trivial on the
$\Z$ component.

{}For $p\in\cP$ define $n^\pm(p)=\sharp(\{j|p_j=\pm\})$, and set
\be
\cP_{l,n-l}=\{p\in\cP|n^+(p)=l,\ n^-(p)=n-l\},\
\cP_n=\cup_{l=0}^n\cP_{l,n-l}.
\en
Note that $c^+_k\cP_{l,n-l}\subset \C[\cP_{l,n-l-1}]$ and
$c^-_k\cP_{l,n-l}\subset \C[\cP_{l-1,n-l}]$.
We also define
\be
\cP_{[k,l]}=\{p\in\cP|p_j=\overline0\ (j\leq k-1),\
p_j=0\ (j\geq l+1)\}.
\en
Our goal is to show
\begin{thm}\label{BASIS}
We assume that $q$ is generic.
There exists an intertwiner of the $\widetilde\cA$ modules
\be
\mathcal X:
\C[\Z\times\cP]\rightarrow\cW_{[\al]}.
\en
\end{thm}
A proof is given in the subsequent subsections.
The uniqueness of the intertwiner is not true.
We need more structures, e.g., creation operators,
in order to single out a unique basis.

Set
$X^{(\al+k)}_p=\mathcal X(k,p)\in\cW_{\al+k}$. The intertwining property  means
\bea
X^{(\beta+1)}_p&=&\kapb(X^{(\beta)}_p),\label{INT0}\\
X^{(\beta)}_{\tau(p)}&=&\taub(X^{(\beta)}_p),\label{INT1}\\
X^{(\beta)}_{s_i(p)}&=&\sb_i(X^{(\beta)}_p),\label{INT2}\\
X^{(\beta)}_{c^\pm_j(p)}&=&\cb^\pm_j(X^{(\beta)}_p).\label{INT3}
\ena
If we define $X^{(\al)}_p$ all other $X^{(\beta)}_p$
are defined by \eqref{INT0}. Our description below exploits this without
mentioning it any further.

Before giving a proof we prepare several statements on the
operators $\cb^\pm_j$.
\begin{prop}
\bea
&&\cb^\pm_j(\alpha)\left(\left(\cW_\alpha\right)_{[k,l]}\right)=0\
\hbox{ unless $k\leq j\leq l$,}\label{ANN1}\\
&&\cb^\pm_l(\alpha)
\left(\left(\cW_\alpha\right)_{[k,l]}\right)\subset
\left(\cW_{\alpha\mp1}\right)_{[k,l-1]}.\label{ANN2}
\ena
\end{prop}
The property \eqref{ANN1} follows immediately from the definition.
The property \eqref{ANN2} follows from Proposition \ref{PROP1}.

\subsection{Vacuum vectors}
Let us consider the common kernel of $\bb_k(\al),\cb_k(\al)$ $(k\in\Z)$.
We call vectors in the kernel {\it vacuum vectors}.
Set $p_{\rm vac}=p^{(1,1,1)}\in\cP_0$.
The annihilation property \eqref{ANN1} implies
\bea
\cb^\pm_k(\al)v^{(\al)}_{p_{\rm vac}}=0.
\label{VAC-SEED}
\ena
We define
\be
X^{(\al)}_{p_{\rm vac}}=v^{(\al)}_{p_{\rm vac}}.
\en
Let $\cW_{\al,{\rm vac}}$ be the subspace of
$\cW_\al$ spanned by the states in the orbit of
$X^{(\al)}_{p_{\rm vac}}$ by ${\bf s}_i$ and $\taub^n$ $(i,n\in\Z)$.
\begin{prop}
{}For generic $q$, the common kernel of the operators
$\bb_k(\al),\cb_k(\al)$ $(k\in\Z)$ is equal to $\cW_{\al,{\rm vac}}$ $:$
\be
\cW_{\al,{\rm vac}}=\bigcap_{k\in\Z}
\left({\rm Ker}\,\bb_k(\alpha)\cap{\rm Ker}\,\cb_k(\alpha)\right).
\en
\end{prop}
\begin{proof}
It is enough to show the equality
\bea
\cW_{\al,{\rm vac}}\cap\left(\cW_\al\right)_{[k,l]}
=\left(\bigcap_{k\in\Z}
\left({\rm Ker}\,\bb_k(\alpha)\cap{\rm Ker}\,\cb_k(\alpha)\right)\right)
\cap\left(\cW_\al\right)_{[k,l]}.
\label{VAC}
\ena
The inclusion $\subset$ is a consequence of \eqref{VAC-SEED} and the equivariance
\eqref{Cov1}, \eqref{Cov2}. We know that for $q=i$
\bea
\dim\cW_{\al,{\rm vac}}\cap
\left(\cW_\al\right)_{[k,l]}=2^{l-k+1}\,,
\label{DIMEQ}
\ena
and \eqref{VAC} holds.
Moreover, for generic $q$, the space
$\cW_{\al,{\rm vac}}\cap\left(\cW_\al\right)_{[k,l]}$
is spanned by the $2^{l-k+1}$ elements in the orbit of
$X^{(\al)}_{p_{\rm vac}}$.
By specialization of the parameter $q$, the dimension of the kernel
does not decrease, while the dimension of the linear span does not increase.
Therefore \eqref{DIMEQ} is valid for generic $q$. The equality \eqref{VAC}
follows from this.
\end{proof}
\subsection{Filtration of $\cW_{[\al]}$}
Proposition \ref{PROP1} suggests a kind of particle structure in
the space $\cW_{[\al]}$ wherein the operators
$\bb_k(\al)$ and $\cb_k(\al)$ act as annihilation operators.
Starting from the subspace $\cW_{\al,{\rm vac}}$ we define
a filtration of $\cW_\al$:
\be
&&0=F^{-1}\cW_\al\subset F^0\cW_\al=\cW_{\al,{\rm vac}}
\subset\cdots\subset F^n\cW_\al\subset\cdots\subset\cW_\al,\\
&&F^n\cW_\al=\sum_{k\in\Z}
\mathfrak S_\infty\left(\cW_\al\right)_{[k,k+n-1]}.
\en
We have
\be
\cb^\pm_k(\al):F^n\cW_\al\rightarrow
F^{n-1}\cW_{\al\mp1}.
\en
Set ${\rm Gr}^n_F\cW_\al=F^n\cW_\al/F^{n-1}\cW_\al$.
We denote the induced mappings by the same symbols:
$\cb^\pm_k(\al):{\rm Gr}^n_F\cW_\al\rightarrow
{\rm Gr}^{n-1}_F\cW_{\al\mp1}$.
We will construct $X^{(\al)}:\C[\cP_n]\rightarrow F^n\cW_\al$
so that the intertwining properties \eqref{INT0}--\eqref{INT3} are satisfied
and the induced mapping
\be
X^{(\al)}:K[\cP_n]\rightarrow{\rm Gr}^n_F\cW_\al
\en
is an isomorphism. Our proof will show that
\be
X^{(\al)}:K[\cP_{[k,l]}]\rightarrow\left(\cW_\al\right)_{[k,l]}
\en
is an isomorphism.
\medskip

{\it Proof of {\rm Theorem \ref{BASIS}}.}\quad
We proceed by induction on $n$. Suppose that we have constructed
$X^{(\al)}$ on $\cup_{j=0}^{n-1}\cP_j$
and the intertwining properties are satisfied.
We construct $X^{(\al)}$ on $\cP_n$.
Set $p^{(l,n-l)}=p^{(1,l+1,n+1)}\in\cP_n$.
The subspace $\C[\cP_{l,n-l}]$ is generated from $p^{(l,n-l)}$ by
the action of $\C[\mathfrak S_\infty]$ and the shift operators $\taub^n$
$(n\in\Z)$. We construct
$X^{(\al)}_{p^{(l,n-l)}}\in(\cW_\al)_{[1,n]}$ which satisfies
\bea
&&{\bf s}_iX^{(\al)}_{p^{(l,n-l)}}=
\begin{cases}
X^{(\al)}_{p^{(l,n-l)}}&\hbox{ if $i< 0$ or $i\geq n+1$;}\\
-X^{(\al)}_{p^{(l,n-l)}}&\hbox{ if $1\leq i\leq l-1$ or $l+1\leq i\leq n-1$.}
\end{cases}\\
\label{INTB1}
&&\cb^\pm_k(\al)X^{(\al)}_{p^{(l,n-l)}}=X^{(\al\mp1)}_{\cb^\pm_kp^{(l,n-l)}}.
\label{INTB2}
\ena
Then, we can induce $X^{(\al)}$ from $p^{(l,n-l)}$
to $\C[\cP_{l,n-l}]$ by using the action
of $\mathfrak S_\infty$ and the shift operators. By construction
the intertwining properties \eqref{INT1} and \eqref{INT2} are satisfied.
Because of \eqref{Cov1} and \eqref{Cov2},
the last one \eqref{INT3} follows from \eqref{INTB2}.

Set $\overline{\cP}_{l,n-l}=\cP_{[1,n]}\cap\cP_{l,n-l}$,
$\overline{\cP}_n=\cup_{l=0}^n\overline{\cP}_{l,n-l}$, and
$\cP^{(k)}=\{p\in\cP|p_k=0\}$.
The intertwining property \eqref{INT3} with \eqref{ACT} implies that if
$p,p'\in\overline{\cP}_{l,n-l}$, we have
\be
\cb^{-p_1}_1(\al+p_2+\cdots+p_n)\cdots\cb^{-p_n}_n(\al)X^{(\al)}_{p'}=
\delta_{p,p'}\varepsilon(p)X^{(\al-2l+n)}_{p_{\rm vac}}\,,
\en
where $\varepsilon(p)$ is given in \eqref{epsilon}. 
Our immediate goal is to construct a family of states
$Y^{(\al)}_p$ $(p\in\overline{\cP}_{l,n-l}\cup(-\overline{\cP}_{l,n-l}))$ 
satisfying
\bea
&&Y^{(\al)}_p=-Y^{(\al)}_{-p},
\label{Y1}
\\
&&Y^{(\al)}_{s_i(p)}=\sb_i(Y^{(\al)}_p),
\label{Y2}
\\
&&\cb^{-p_1}_1(\al+p_2+\cdots+p_n)\cdots\cb^{-p_n}_n(\al)Y^{(\al)}_{p'}=
\delta_{p,p'}\varepsilon(p)X^{(\al-2l+n)}_{p_{\rm vac}}.
\label{Y3}
\ena
{}For each $0\leq l\leq n$ take an $\mathfrak S_{[1,n]}$ invariant
subspace $\mathcal U_{l,n-l}\subset F^n\cW_\al\cap\cW_\al^{(2l-n)}$ of dimension
\be
\sharp(\overline{\cP}_{l,n-l})=\begin{pmatrix}n\\l\end{pmatrix}
\en
such that $\oplus_{l=0}^n\mathcal U_{l,n-l}\rightarrow{\rm Gr}^n_F\cW_\al$
is an isomorphism. The non-uniqueness of such spaces is the reason
for the non-uniqueness of the basis. We have no a priori reason to choose one.

{}Fix an arbitrary basis of $\mathcal U_{l,n-l}$,
$\{v_p|p\in\overline{\cP}_{l,n-l}\}$.
Consider the matrix $M=(M_{p,p'})_{p,p'\in\overline{\cP}_{l,n-l}}$
given by
\be
\cb^{-p_1}_1(\al+p_2+\cdots+p_n)\cdots\cb^{-p_n}_n(\al)v_{p'}=
M_{p,p'}X^{(\al-2l+n)}_{p_{\rm vac}}.
\en
{}From Proposition \ref{PROP1} we see that the left hand side
is proportional to $X^{(\al-2l+n)}_{p_{\rm vac}}$.
The matrix $M$ is invertible for generic $q$ because it is so for
the fermion case $q=i$. Therefore, there exists a unique set of elements
$Y^{(\al)}_p\in\mathcal U_{l,n-l}$
$(p\in\overline{\cP}_{l,n-l}\cap(-\overline{\cP}_{l,n-l}))$ where
$Y^{(\al)}_{-p}=-Y^{(\al)}_p$ satisfying \eqref{Y3}
for $p,p'\in\overline{\cP}_{l,n-l}$.
The equivariance \eqref{Y2} follows from \eqref{Cov1}, the Grassmann relation
and the uniqueness of the solution to the linear equation \eqref{Y3}.

We will modify $Y^{(\al)}_p$ to $X^{(\al)}_p$
by adding lower order terms in the filtration.
Take $p\in\overline{\cP}_{l,n-l}$.
For each $k$, consider the state
\be
\cb^\pm_k(\al)Y^{(\al)}_p\in
{\bf s}_k\cdots{\bf s}_{n-1}\left(\left(\cW_\al\right)_{[1,n-1]}\right)\,.
\en
By the induction hypothesis, it can be written as
\be
\cb^\pm_k(\al)Y^{(\al)}_p=\sum_{p'\in\cP^{(k)}\cap\cP_{[1,n]}}
f^\pm_{k,p,p'}X^{(\al\mp1)}_{p'}.
\en
If $p'\in\cP_{n-1}$ appears in the sum,
we have $p_k=\mp$, $p'=p[k]$ and $f^\pm_{k,p,p'}=\varepsilon^\mp(p,k)$. 
For, if $p_k=\pm$, or $p_k=\mp$ but $p'$ is an element of $\cP_{n-1}$ other than
$p[k]$, $Y^{(\al)}_p$ can be brought to non-zero multiple of
$X^{(\al-2l+n)}_{p_{\rm vac}}$ by a sequence of operators other than
$\cb^{-p_1}_1(\al+p_2+\cdots+p_n)\cdots\cb^{-p_n}_n(\al)$. 
This is a contradiction to \eqref{Y3}. The equality
$f^\pm_{k,p,p[k]}=\varepsilon^\mp(p,k)$ follows from
\be
&&\cb^{-p_1}_1(\al+p_2+\cdots+p_n)\cdots\cb^{-p_n}_n(\al)Y^{(\al)}_p\\
&&=(-1)^{n-k}f^\pm_{k,p,p[k]}
\cb^{-p_1}_1(\al+p_2+\cdots+p_n)\cdots\buildrel k\over\vee
\cdots\cb^{-p_n}_n(\al)X^{(\al\mp1)}_{p[k]}.
\en
Note that $\varepsilon^\mp(p,k)p[k]=c^\pm_k(p)$.

If $p'\not\in\cP_{n-1}$
appears in the sum, the total spin conservation requires
\bea
p'\in
\begin{cases}
\cP_{l-i,n-l-1-i}\cap\cP^{(k)}\cap\cP_{[1,n]}&\hbox{ for $\cb^+_k(\al)$;}\\
\cP_{l-1-i,n-l-i}\cap\cP^{(k)}\cap\cP_{[1,n]}&\hbox{ for $\cb^-_k(\al)$.}
\end{cases}\label{SUM}
\ena
Here $i\geq1$. Therefore, we have
\be
\cb^\pm_k(\al)Y^{(\al)}_p=X^{(\al\mp1)}_{\cb^\pm_kp}
+\sum_{p'}f^\pm_{k,p,p'}X^{(\al\mp1)}_{p'}
\en
where the sum over $p'$ is restricted as \eqref{SUM}.

Set
\be
X^{(\al)}_p=Y^{(\al)}_p-\sum_{\tilde p\in\cP_{[1,n]}
\cap(\cP_{n-2}\cup\cP_{n-4}\cup\cdots)}g_{\tilde p}X^{(\al)}_{\tilde p}.
\en
We do not sum over $\tilde p\in\cP_0$.
We require $\cb^\pm_k(X^{(\al)}_p)=X^{(\al\mp1)}_{\cb^\pm_kp}$.
This is equivalent to
\be
g_{\tilde p}
=f^{-\tilde p_k}_{k,p,\tilde p[k]}\,\varepsilon^{\tilde p_k}(\tilde p,k)
\en
for all $1\leq k\leq n$ such that $\tilde p_k=\pm$.
The Grassmann relation implies that the right hand side is independent of $k$:
suppose that $\tilde p_k,\tilde p_l\not=0,\overline0$.
{}From
\be
\cb^{-\tilde p_l}_l(\al+\tilde p_k)\cb^{-\tilde p_k}_k(\al)Y^{(\al)}_p
=-\cb^{-\tilde p_k}_k(\al+\tilde p_l)\cb^{-\tilde p_l}_l(\al)Y^{(\al)}_p
\en
we obtain
$f^{-\tilde p_k}_{k,p,\tilde p[k]}\varepsilon^{\tilde p_l}(\tilde p[k],l)
=-f^{-\tilde p_l}_{l,p,\tilde p[l]}\varepsilon^{\tilde p_k}(\tilde p[l],k)$.
{}From
\be
\cb^{-\tilde p_l}_l(\al+\tilde p_k)
\cb^{-\tilde p_k}_k(\al)X^{(\al)}_{\tilde p}
=-\cb^{-\tilde p_k}_k(\al+\tilde p_l)
\cb^{-\tilde p_l}_l(\al)X^{(\al)}_{\tilde p}
\en
we obtain
$\varepsilon^{\tilde p_k}(\tilde p,k)
\varepsilon^{\tilde p_l}(\tilde p[k],l)
=-\varepsilon^{\tilde p_l}(\tilde p,l)
\varepsilon^{\tilde p_k}(\tilde p[l],k)$.

Since $Y^{(\al)}_p$ is equivariant by \eqref{Y2},  
and the construction of
$X^{(\al)}_p$ is equivariant, we have
\be
{\bf s}_i(X^{(\al)}_p)=X^{(\al)}_{s_i(p)}\ (1\leq i\leq n-1).
\en
\qed

\appendix

\section{Anti-commutativity of $\mathbf{c}(\z,\al)$}\label{sec:app1}

In this section we sketch the derivation for 
the anti-commutativity property of the operators $\cb(\z,\al)$.  
Anti-commutativity of $\cb(\z,\al)$ with $\bb(\z,\al)$    
requires a different reasoning which is given in \cite{BJMST}. 
We do not repeat it here. 

We need information about the $R$ matrix. 
First assume that $q$ is not a root of unity. 
Define $R_{A,B}(\z)\in\End(W^+_A\otimes W^+_B)$ by
\be
&&R_{A,B}(\z)=P_{A,B}h(\z,u_{A,B})\z^{D_A+D_B}, 
\\
&&u_{A,B}=\ao_A^*q^{-2D_A}\ao_B\,,
\en
where $P_{A,B}$ denotes the permutation, and  
$h(\z,u)$ is the unique formal power series in $u$ satisfying 
\bea
&&h(\z,q^2u)(1+\z^{-1} u)=h(\z,u)(1+\z u)\,,
\label{eq:hu}
\\
&&h(\z,0)=1\,.
\nn
\ena
When there exists a positive integer $r$ such that
$q^{2r}=1$ and $q^{2j}\neq 1$ ($1\le j\le r-1$), 
the representation $W^+_A$ has an $r$-dimensional quotient
$W^+_{A,0}$ generated by $|0\rangle$. 
In this case $u_{A,B}^{r}=0$. 
Define $R_{A,B}(\z)\in\End(W^+_{A,0}\otimes W^+_{B,0})$ 
by the same formula, where $h(\z,u)$ is the unique element 
in the truncated polynomial ring 
$\C[u]/(u^r)$ with the above properties. 

\begin{lem}
The operator $R_{A,B}(\z)$ satisfies the intertwining property
\be
&&R_{A,B}(\z_1/\z_2)
L^+_{A,j}(\z_1)L^+_{B,j}(\z_2)
=
L^+_{B,j}(\z_2)L^+_{A,j}(\z_1)
R_{A,B}(\z_1/\z_2)\,.
\en
\end{lem}
\begin{proof}
If $uv=vu+(1-q^2)w$ and $uw=q^{-2}wu$, then 
\be
f(u)v=vf(u)+\frac{f(u)-f(q^2u)}{u} w
\en
holds for any $f(u)$. 
Taking $u=u_{A,B}$ and  
$v=\ao^*_Bq^{2D_A}$, $w=\ao^*_A$
or $v=\ao_A$, $w=-\ao_Bq^{-2D_A}$, we deduce from 
\eqref{eq:hu} the identities
\be
&&h(\z,u)(\z \ao^*_A+\ao^*_Bq^{2D_A})=
(\z^{-1}\ao^*_A+\ao^*_Bq^{2D_A})h(\z,u),
\\
&&h(\z,u)(\ao_A+\z^{-1}\ao_Bq^{-2D_A})=
(\ao_A+\z \ao_Bq^{-2D_A})h(\z,u)\,.
\en
The intertwining relations follow from these. 
\end{proof}

The $R$ matrix which intertwines the fused L operators
\bea
&&R_{\{A,a\},\{B,b\}}(\z_1/\z_2)
L^+_{\{A,a\},j}(\z_1)L^+_{\{B,b\},j}(\z_2)
\label{RLLbig}\\
&&\quad
=
L^+_{\{B,b\},j}(\z_2)L^+_{\{A,a\},j}(\z_1)
R_{\{A,a\},\{B,b\}}(\z_1/\z_2)\,
\nn
\ena
has the triangular form 
\be
R_{\{A,a\},\{B,b\}}(\z)
=
\begin{pmatrix}
R_{11} & 0 & 0 & 0 \\
R_{21}&R_{22} &0 &0\\
R_{31}&0 &R_{33} &0\\
R_{41}&R_{42} &R_{43} &R_{44}\\
\end{pmatrix}\,.
\en
Its non-zero entries read as follows. 
\be
&&R_{11}=q^{D_B}R_{A,B}(\z)q^{-D_A},
\\
&&R_{21}=-q^{-1}\z^2 
q^{D_A-D_B}R_{A,B}(q^{-1}\z)\ao_B
\\
&&R_{22}=q(1-q^{-2}\z^2)q^{D_A}R_{A,B}(q^{-2}\z) q^{D_B},
\\
&&R_{31}=-\frac{\z}{1-q^2\z^2} q^{D_B-D_A}
R_{A,B}(\z q)\ao_B, 
\\
&&R_{41}=\frac{q\z^3}{1-q^2\z^2}
q^{-D_B}R_{A,B}(\z)q^{-D_A} \ao_B^2,
\\
&&R_{42}=-q\z\cdot
R_{A,B}(\z q^{-1})q^{-D_A+D_B}
\ao_B,
\\
&&R_{33}=-\frac{q\z^2}{1-q^2\z^2}
q^{-D_A}R_{A,B}(q^{2}\z) q^{-D_B},
\\
&&R_{43}=\frac{q^2\z^4}{1-q^2\z^2}
q^{D_A-D_B}R_{A,B}(\z q)q^{-2D_B}\ao_B,
\\
&&R_{44}=-\z^2 q^{D_A}R_{A,B}(\z)q^{-D_B}.
\en
The anti-commutativity of $\cb(\z,\al)$ is an immediate consequence of the following Lemma. 

\begin{lem} We have
\be
&&
\bk^+(\z_1,\al-1)\bk^+(\z_2,\al)
+\bk^+(\z_2,\al-1)\bk^+(\z_1,\al)
\\
&&=
\Delta_{q,\z_1}F(\z_1,\z_2)+\Delta_{q,\z_2}F(\z_2,\z_1)\,,
\en
where
\be
&&F(\z_1,\z_2)
=q^{\al-1}\frac{1}{1-\z_1^2/\z_2^2}\bQ^{(1)}(\z_1,\al-1)
\bk^{(1)}(\z_2,\al)
\\
&&\quad+
q^{\al-1}\frac{\z_1/\z_2}{1-\z_1^2/\z_2^2}\bQ^{(2)}(\z_1,\al-1)
\bk^{(0)}(\z_2,\al-1)
\\
&&\quad
-q^{2(\al-1)}
\frac{\z_1/\z_2}{1-\z_1^2/\z_2^2}\bQ^{(2)}(\z_1,\al-1)
\Delta_{q,\z_2}\bQ^{(0)}(\z_2,\al-1)\,,
\en
and
\be
&&
\bQ^{(k)}(\z,\al)(X)
=(1-q^{2(\al-\bS)})
\z^{\al-\bS}
{\rm Tr}^+_A\bigl(q^{2\al D_A}\ao_A^k
T^+_A(\z)^{-1}\,X\,T^+_A(\z)\bigr)\,,
\\
&&\bk^{(k)}(\z,\al)(X)
=(1-q^{2(\al-\bS)})\z^{\al-\bS}
{\rm Tr}^+_A{\rm tr}_a\left(
q^{2\al D_A}(\ao_A^*)^{1-k}\sigma^+_a
T^+_{\{a,A\}}(\z)^{-1} \,X\,T^+_{\{a,A\}}(\z)\right)\,.
\en
\end{lem}
\begin{proof}
Set 
\be
(T^+_{\{A,a\}}(\z))^{-1} X T^+_{\{A,a\}}(\z)=
\begin{pmatrix}
\cA(\z)(X)& 0 \\
\cC(\z)(X)&  \cD(\z)(X) \\
\end{pmatrix}\,.
\en
For brevity we write $\z=\z_1/\z_2$, $R_{ij}=R_{ij}(\z)$, 
$\cA_1=\cA(\z_1)$, and so forth. 
The intertwining relation \eqref{RLLbig} contains the following relations.
\be
&&R_{21}\cA_2\cA_1+R_{22}\cC_2\cA_1=\cA_1\cC_2R_{11}+\cA_1\cD_2R_{21},
\\
&&R_{42}\cD_2\cA_1+R_{44}\cD_2\cC_1=\cC_1\cD_2R_{22}+\cD_1\cD_2R_{42},
\\
&&R_{41}\cA_2\cA_1+R_{42}\cC_2\cA_1+R_{43}\cA_2\cC_1+R_{44}\cC_2\cC_1
\\
&&\quad=\cC_1\cC_2R_{11}+\cC_1\cD_2R_{21}+\cD_1\cC_2R_{31}+\cD_1\cD_2R_{41}\,.
\en
Eliminating $\cC_2\cA_1$ and $\cC_1\cD_2$, we obtain
\bea
&&\cC_1\cC_2-R_{44}\cC_2\cC_1R_{11}^{-1}
=X\cA_1\cA_2-\cD_1\cD_2X
\label{**}\\
&&\qquad +R_{43}\cA_2\cC_1R_{11}^{-1}
-R_{44}\cD_2\cC_1 R_{22}^{-1}R_{21}R_{11}^{-1}
+R_{42}R_{22}^{-1}\cA_1\cC_2-\cD_1\cC_2 R_{31}R_{11}^{-1}\,,
\nn
\ena
where $X=(R_{41}-R_{42}R_{22}^{-1}R_{21})R_{11}^{-1}$. 
From the explicit form of the matrix elements we calculate
\be
&&X=\left(\frac{q\z}{1-q^2\z^2}-\frac{q^{-1}\z}{1-q^{-2}\z^2}\right)a_A^2q^{-2D_B},
\\
&&R_{11}^{-1}q^{-2D_A}R_{43}=\frac{q^4\z^4}{1-q^2\z^2}(1+q^{-1}\z^{-1}u_{A,B})a_Bq^{-2D_B},
\\
&&R_{22}^{-1}R_{21}R_{11}^{-1}q^{-2D_A}R_{44}=
\frac{q^{-2}\z^4}{1-q^{-2}\z^2}(1+q\z^{-1}u_{A,B})a_Bq^{-2D_B},
\\
&&q^{-2D_A}R_{42}R_{22}^{-1}=-\frac{q^2}{1-q^{-2}\z^2}(1+q^{-1}\z u_{A,B})a_Aq^{-2D_A},
\\
&&R_{31}R_{11}^{-1}
=-\frac{1}{1-q^2\z^2}(1+q\z u_{A,B})a_A.
\en
Multiply both sides of \eqref{**} by 
\be
(1-q^{2(\al-\bS-1)})(1-q^{2(\al-\bS+1)})
\z_1^{\al-\bS-1}\z_2^{\al-\bS+1}
q^{2(\al-1)D_A+2\al D_B} 
\en
and take the trace. 
Direct calculation leads to the assertion. 
\end{proof}

\section{Proof of Theorem \ref{thm:tQ},\ref{thm:c}}\label{sec:app2}

In this section, we give a derivation of Theorems 
\ref{thm:tQ},\ref{thm:c}.

\subsection{Preliminaries}

Throughout this section, 
we fix $n$ and work with the interval $[1,n]$. 
To perform the calculation we find it technically 
easier to pass from $\cV=\End(V^{\otimes n})$ 
to the $2n$ fold tensor product of $V$.  
Let $\iota:\cV\overset{\sim}{\rightarrow} V^{\otimes 2n}$ 
denote the isomorphism of vector spaces given by 
\be
\iota\left(E_{\e_1,\e_1'}\otimes\cdots\otimes E_{\e_n,\e_n'}\right)
=\prod_{j=1}^n\e_j'\cdot
v_{\e_1}\otimes\cdots\otimes v_{\e_n}
\otimes v_{-\e_n'}\otimes\cdots\otimes v_{-\e_1'},
\en
where $E_{\e,\e'}=\bigl(\delta_{\e\mu}\delta_{\e'\nu}\bigr)_{\mu,\nu}$ 
stands for the matrix unit.
In what follows, we set
\be
\bar{k}=2n+1-k,\quad \xi_{\bar{k}}=q^{-1}\xi_k\,
\quad (k=1,\cdots,n).  
\en
Under the isomorphism $\iota$, 
the left and the right multiplication by an element $Z_k$ 
($Z\in\End(V)$) are translated respectively into 
\be
\iota(Z_kX)=Z_k\cdot\iota(X),
\qquad
\iota(XZ_k)=\bigl(\sigma^2\ {}^tZ\ \sigma^2\bigr)_{\bar{k}}\cdot
\iota(X).
\en
In particular we have
$\iota\bigl(\bS(X)\bigr)=\widetilde{S}\cdot\iota(X)$ and
$\iota(X\,S)=-\overline{S}\cdot\iota(X)$,  
where
\be
\widetilde{S}=\frac{1}{2}\sum_{j=1}^{2n}\sigma^3_k,
\quad
\overline{S}=\frac{1}{2}\sum_{k=1}^n\sigma^3_{\bar k}.
\en
We shall also make use of the relations
\be
&&\iota\left(X\ L_{A,j}^\pm(\z)\right)
=(q\z-q^{-1}\z^{-1})L^\pm_{A,\bar{j}}(q\z)^{-1}\cdot
\iota(X),
\\
&&\iota\left(X{L_{A,j}^\pm(\z)}^{-1}\right)
=(\z-\z^{-1})^{-1}L^\pm_{A,\bar{j}}(q^{-1}\z)\cdot
\iota(X)\,,
\en
which hold provided the entries of $X$ commute with $Osc$. 

We begin by rewriting Q operators.  Introduce operators
\be
&&
\psit^\pm_j=\sigma^\pm_jq^{\pm\sum_{l=j+1}^{2n}\sigma^3_l}\quad
(j=1,\cdots,2n),
\\
&&
\psih^\pm_j=\psit^\pm_jq^{\mp2\overline{S}},
\quad
\psih^\pm_{\bar j}=\psit^\pm_{\bar j}
q^{\pm2(\widetilde{S}-\overline{S})},
\quad (j=1,\cdots,n),
\en
and set 
\be
&&\cL_{A,j}(\z)=1+\z\ao^*\psit^+_j+\z\ao\ \psit^-_j
-\z^2q^{2D}\psit^+_j\psit^-_j\,,
\\
&&\cLh_{A,j}(\z)=
1-\z\ao^*\psih^+_j-\z\ao\ \psih^-_j
-\z^2q^{2D+2}\psih^-_j\psih^+_j\,.
\en

\begin{lem}\label{lem:QQ*}
{}For $X\in \cV^{(s)}$, the following hold.
\bea
&&\iota\bigl(\z^{-\al+\bS}
\bQ^+(\z,\al)(X)\bigr)
\label{bQ1}\\
&&\quad=
\frac{1-q^{2(\al-s)}}{\prod_{l=1}^n(1-\z^2/\xi_l^2)}
\Tr^+_A\left(
q^{2(\al-s)D}\cL_{A,1}(\z/\xi_1)\cdots\cL_{A,2n}(\z/\xi_{2n})\right)
\cdot\iota(X)\,,
\nn
\\
&&\iota\bigl(\z^{\al-\bS}
\bQ^{*-}(\z,\al)(X)\bigr)
\label{bQ2}\\
&&\quad=-
\frac{1-q^{-2(\al-s)}}{\prod_{l=1}^n(1-\z^2/\xi_l^2)}
\Tr^-_A\bigl(
\cLh_{A,n}(\z/\xi_n)\cdots\cLh_{A,1}(\z/\xi_1)
\nn\\
&&\quad\times
q^{-2\al D_A}
\cLh_{A,\bar 1}(\z/q\xi_1)\cdots\cLh_{A,\bar n}(\z/q\xi_n)
q^{2s D_A}\bigr)\cdot\iota(X)\,.
\nn
\ena
\end{lem}
\begin{proof}
This can be shown by direct calculation noting that 
\be
&&
i\z^{-1/2}q^{1/4}\cL_{A,j}(\z)
=q^{\sum_{l=j}^{2n}\sigma^3_l D}\cdot 
(\z-\z^{-1})L^+_{A,j}(\z)^{-1}\cdot
q^{-\sum_{l=j+1}^{2n}\sigma^3_l D}\,,
\\
&&i\z^{-1/2}q^{-1/4}\cLh_{A,j}(\z)
\\
&&\quad=
\begin{cases}
q^{\sum_{l=j+1}^{n}\sigma^3_l \cdot D_A}
\cdot L^+_{A,j}(\z)\cdot
q^{-\sum_{l=j}^{n}\sigma^3_l \cdot D_A}
& (1\le j\le n),\\
q^{(\sum_{l=j+1}^{2n}\sigma^3_l+\sum_{l=1}^{n}\sigma^3_l) D_A}
\cdot L^+_{A,j}(\z)\cdot
q^{-(\sum_{l=j}^{2n}\sigma^3_l+\sum_{l=1}^{n}\sigma^3_l)D_A}
& (n+1\le j\le 2n).\\
\end{cases}
\en
\end{proof}

This rewriting is useful at $q=i$, when 
$\psit^\pm_j$, $\psih^\pm_j$ become the Jordan-Wigner 
fermions on $V^{\otimes 2n}$. 
In the rest of this section we shall consider only this case.

\subsection{Diagonalization of Q operators}

Let us calculate the trace \eqref{bQ1}. 
{}First note the following simple fact. 

\begin{lem}\label{lem:rootof1}
If $q$ is a primitive $r$-th root of $1$, the representation
$W^+$ of $Osc$ has an $r'$-dimensional 
quotient $W^{+}_0$ generated by $\ket{0}$, where
\be
r'=\begin{cases}
r & \text{($r$ odd)}, \\
r/2 & \text{($r$ even)}. \\
\end{cases}
\en
We have the relation 
\be
\Tr^+\bigl(y^{2D_A}x\bigr)
=\frac{1}{1-y^{2 r'}}{\rm tr}_{W^{+}_0}
\bigl(y^{2D_A}x\bigr) 
\qquad (x\in Osc).
\en
\end{lem}

When $q=i$, $W^+_0$ is two-dimensional with basis $|0\rangle,|1\rangle$. 
The $q$-oscillators are represented in this basis as 
\be
a\mapsto \begin{pmatrix} 0 & 2\\ 0 & 0\\ \end{pmatrix},
\quad 	
a^*\mapsto \begin{pmatrix} 0 & 0\\ 1 & 0\\ \end{pmatrix},
\quad
i^{2D_A}\mapsto 
\begin{pmatrix} 1 & 0\\ 0 & -1\\ \end{pmatrix}, 
\en
while
\be
\cL_{A,k}(\z)\mapsto
\begin{pmatrix}
1-\z^2\psit^+_k \psit^-_k & 2\z\psit^-_k\\
\z\psit^+_k & 1+\z^2\psit^+_k \psit^-_k
\end{pmatrix} \,.
\en

\begin{lem}\label{lem:Gras}
Let $\eta_j,\eta_j^*$ ($j=1,\cdots,N$) be generators of 
a Grassmann algebra, and set  
\be
L_j=
\begin{pmatrix}
1+\eta_j^*\eta_j/2 & \eta_j^* \\
\eta_j &1-\eta_j^*\eta_j/2 \\
\end{pmatrix}, 
\quad 
H=
\begin{pmatrix} 1 & 0\\ 0 & t \\ \end{pmatrix}\,.
\en
Then 
\bea
&&{\rm tr}(L_N\cdots L_1 H)
=
(1+t)\exp\bigl(-\frac{t}{1+t}\sum_{j,k=1}^N\eta_k^*\eta_j
+\frac{1}{2}\sum_{j=1}^N\eta^*_j\eta_j
+\sum_{j<k}\eta^*_k\eta_j\bigr). 
\label{eq:lem1}
\ena
\end{lem}
\begin{proof}
{}First consider the case $t=0$.  
Extracting the factor 
$e^{\frac{1}{2}\sum_{j=1}^N\eta^*_j\eta_j}$, 
we are to show that 
\bea
{\rm tr}\left(L'_N \cdots L'_1 
\begin{pmatrix} 1 & 0\\ 0 & 0 \\ \end{pmatrix}\right)
=e^{\sum_{j<k}\eta^*_k\eta_j} 
\label{eq:KN}
\ena
where 
$L'_j=\begin{pmatrix}1 & \eta_j^* \\
\eta_j &1-\eta_j^*\eta_j \\\end{pmatrix}$. 
Denote the left hand side of \eqref{eq:KN} by $x_N$. 
Using 
\be
\begin{pmatrix}
1 & 0 \\
0 & 0 \\
\end{pmatrix} 
L'_N=
\begin{pmatrix}
1 & 0 \\
0 & 0 \\
\end{pmatrix} 
\begin{pmatrix}
1 & \eta_N^* \\
0 &1 \\
\end{pmatrix}, 
\quad
\begin{pmatrix}
1 & \eta_N^* \\
0 &1 \\
\end{pmatrix} 
L'_j
\begin{pmatrix}
1 & -\eta_N^* \\
0 &1 \\
\end{pmatrix} 
=e^{\eta^*_N\eta_j}L'_j,
\en
we find a recurrence relation 
$x_N=e^{\eta^*_N\sum_{j=1}^{N-1}\eta_j}x_{N-1}$.  
Eq. \eqref{eq:KN} follows from this. 

In general, \eqref{eq:lem1} is linear in $t$ and 
the coefficient of $t$ is obtained from $t=0$ 
by exchanging the roles of $\eta^*_j$ and $\eta_j$. 
Combining them we obtain 
\be
&&
\exp\left(
\frac{1}{2}\sum_{j=1}^N\eta^*_j\eta_j
+\sum_{j<k}\eta^*_k\eta_j\right)
+t 
\exp\left(\frac{1}{2}\sum_{j=1}^N\eta_j\eta^*_j
+\sum_{j<k}\eta_k\eta^*_j\right)
\\
&&=\left(1+t(1-\sum_{j,k=1}^N\eta^*_k\eta_j)\right)
\exp\left(\frac{1}{2}\sum_{j=1}^N\eta^*_j\eta_j
+\sum_{j<k}\eta^*_k\eta_j\right) 
\\
&&=
(1+t)\exp\left(
 -\frac{t}{1+t}\sum_{j,k=1}^N\eta_k^*\eta_j\right)
\exp\left(\frac{1}{2}\sum_{j=1}^N\eta^*_j\eta_j
+\sum_{j<k}\eta^*_k\eta_j\right).
\en
Lemma is proved. 
\end{proof}
Now set
\be
&&\phi^\pm_k=-i\psit^\pm_k+\psit^\pm_{\bar{k}},
\\
&&\phi^{*\pm}_k(\al)=\frac{1}{i^{\pm \al}-i^{\mp \al}}
\left(i^{\pm\al+1}\psit^\pm_k-i^{\mp \al}\psit^\pm_{\bar{k}}\right)\,.
\en
They are related to the fermions $\Psi^\pm_k,\Phi^\pm_k(\al)$ via
\be
&&\phi^\pm_k\iota(X)=(-1)^s\iota\left(\Psi^\pm_k(X)\right),
\\
&&
\phi^{*\pm}_k(\al)\iota(X)=(-1)^{s+1}
\iota\left(\Phi^\pm_k(\al+s-1)(X)\right)\,,
\en
where $X\in\cV^{(s)}$.  
\begin{lem}\label{lem:Qt}
We have 
\bea
&&\frac{1-i^{2\al}}{\prod_{l=1}^n(1-\z^2/\xi_l^2)}
\Tr^+_A\left(i^{2\al D_A}\cL_{A,1}(\z/\xi_1)\cdots\cL_{A,2n}(\z/\xi_{2n})
\right)=
\nn
\\
&&
\exp
\bigl[
\sum_{j,k=1}^n
\bigl(\log(1+\z^2 M)_{jk}\phi^{*+}_j(\al+1)\phi^-_k
 -\log(1-\z^2 M)_{jk}\phi^{*-}_j(\al+1)\phi^+_k\bigr)
\bigr],
\nn
\ena
where $M$ is an upper triangular matrix with entries
\be
&&
M_{jk}=
\begin{cases}
2\xi_j^{-1}\xi_k^{-1} & (j<k),\\
\xi_j^{-2} & (j=k), \\
0 & (j>k). \\
\end{cases}
\en
\end{lem}
\begin{proof}
We compute the trace first in the normal-ordered form, 
where normal ordering means that 
we bring all $\psit^+_k$ to the left and $\psit^-_k$ to the right. 
Taking $N=2n$, $t=i^{2\al}$ and  
\be
&&\eta^*_k=\frac{2\z}{\xi_k}\psit^-_k,
\quad 
\eta_k=\frac{\z}{\xi_k}\psit^+_k,
\en
we apply formula \eqref{eq:lem1} under the normal ordering symbol $:~:$.
The result is 
\be
&&(1-i^{2\al})\Tr^+_A\left(i^{2\al D_A}
\cL_{A,1}(\z/\xi_1)\cdots\cL_{A,2n}(\z/\xi_{2n})\right)
\\
&&=
:\exp\sum_{j,k=1}^n\z^2M_{jk}
\left(\phi^{*+}_j(\al+1)\phi^-_k-\phi^+_k\phi^{*-}_j(\al+1)\right):\,.
\en   
Due to the formula
\be
&&:\exp\Bigl(\sum_{j,k=1}^{2n}A_{jk}\psit^+_j\psit^-_k\Bigr)\!\!:~
=\exp\left(\sum_{j,k=1}^{2n}\bigl(\log(I+A)\bigr)_{jk}
\psit^+_j\psit^-_k\right)\,, 
\en
the right hand side is rewritten as 
\be
&&
\exp\left(\sum_{j,k=1}^n
\bigl(\log(1+\z^2 M)_{jk}\phi^{*+}_j(\al+1)\phi^-_k
+\log(1-\z^2 M)_{jk}\phi^+_k\phi^{*-}_j(\al+1)\bigr)\right)
\\
&&
=\det(1-\z^2 M) 
\\
&&\quad\times
\exp\left(\sum_{j,k=1}^n
\bigl(\log(1+\z^2 M)_{jk}\phi^{*+}_j(\al+1)\phi^-_k
 -\log(1-\z^2 M)_{jk}\phi^{*-}_j(\al+1)\phi^+_k\bigr)\right)\,.
\en
Lemma follows from this. 
\end{proof}

\begin{lem}\label{lem:Qt*}
Set $\mathcal{T}=\cLh_{A,n}(\z/\xi_n)\cdots\cLh_{A,1}(\z/\xi_1)$ and 
$\overline{\mathcal{T}}=\cLh_{A,\bar{1}}(\z/i\xi_1)\cdots\cLh_{A,\bar{n}}(\z/i\xi_{n})$. 
Then we have on $\iota(\cV^{(s)})$ 
\bea
&&-\frac{1-i^{-2(\al-s)}}{\prod_{l=1}^n(1-\z^2/\xi_l^2)}
i^{2\overline{S}}\cdot
\Tr^-_A\left(
\mathcal{T}\,i^{-2\al D_A}\,\overline{\mathcal{T}}\,
i^{2s D_A}\right)
\cdot i^{-2\overline{S}}
\nn
\\
&&
=\exp
\bigl[
\sum_{j,k=1}^n
\bigl(\log(1+\z^2 M)_{jk}\phi^-_k\phi^{*+}_j(\al-s+1)
 -\log(1-\z^2 M)_{jk}\phi^+_k\phi^{*-}_j(\al-s+1)\bigr)
\bigr].
\nn
\ena
\end{lem}
\begin{proof}
Let $\mathcal{T}_{even}$ (resp. $\mathcal{T}_{odd}$) be the sum of terms in $\mathcal{T}$ 
containing an even (resp. odd) number of fermions, and similarly for 
$\mathcal{\overline{T}}$. Since the total spin is preserved, we have
\be
\Tr^-_A\left(\mathcal{T}\,i^{-2\al D_A}\,\overline{\mathcal{T}}\,i^{2s D_A}\right)
&=&
\Tr^-_A\left(\mathcal{T}_{even}i^{-2\al D_A}\overline{\mathcal{T}}_{even}i^{2s D_A}\right)
+\Tr^-_A\left(\mathcal{T}_{odd}i^{-2\al D_A}\overline{\mathcal{T}}_{odd}i^{2s D_A}\right)
\\
&=&
\Tr^-_A\left(\bigl(\overline{\mathcal{T}}_{even}-
\overline{\mathcal{T}}_{odd}\bigr)i^{2s D_A}\mathcal{T}i^{-2\al D_A}\right)
\\
&=&
i^{2\overline{S}}\cdot
\Tr^-_A\left(\overline{\mathcal{T}}\,
i^{2s D_A}\,\mathcal{T}\,i^{-2\al D_A}\right)
\cdot i^{-2\overline{S}}.
\en
By using $\cLh_{A,\bar{j}}(\z/i\xi_j)=
i^{-\sigma^3_{\bar{j}}}\cLh_{A,\bar{j}}(i\z/\xi_j)
i^{\sigma^3_{\bar{j}}}$, 
the right hand side can be rewritten further as 
\be
&&i^{2s\overline{S}}
\Tr^-_A\bigl(
i^{-2(\al-s)D_A}\cLh_{A,\bar{1}}(i\z/\xi_1)\cdots\cLh_{A,\bar{n}}(i\z/\xi_{n})
\cLh_{A,n}(\z/\xi_n)\cdots\cLh_{A,1}(\z/\xi_1)
\bigr)i^{-2s\overline{S}}\,.
\nn
\en
The action of $Osc$ on $W^+_0$ factors through 
that of the quotient algebra $Osc_0$
by the relation $i^{4D}=1$.  
Let $\theta$ be the anti-automorphism of the latter
given by 
$\theta(a)=-a$, $\theta(a^*)=-a^*$, $\theta(i^{D})=i^{1-D}$. 
Denote also by $\theta$ the anti-algebra map of Clifford algebras sending 
$\psit^\pm_j$ to $\psih^\pm_j$ ($j=1,\cdots,2n$).  
It is easy to check that 
\be
&&\theta(\cL_{A,j}(\z))=\cLh_{A,j}(\z),
\\
&&
\theta\bigl(
(1-i^{2\al}){\rm Tr}^+_A(i^{2\al D_A}x)
\bigr)
=-(1-i^{-2\al}){\rm Tr}^-_A(i^{-2\al D_A}\theta(x))\,,
\en
where $x\in Osc_0$.
Using these we obtain 
\be
&&-\frac{1-i^{-2(\al-s)}}{\prod_{l=1}^n(1-\z^2/\xi_l^2)}
i^{2\overline{S}}\cdot
\Tr^-_A\left(
\mathcal{T}\,i^{-2\al D_A}\,\overline{\mathcal{T}}\,
i^{2s D_A}\right)
\cdot i^{-2\overline{S}}
\\
&&=
i^{2(s+1)\overline{S}}\theta
\Bigl(
\frac{1-i^{2(\al-s)}}{\prod_{l=1}^n(1-\z^2/\xi_l^2)}
\Tr^+_A\left(i^{2(\al-s)D_A}
\cL_{A,1}(\z/\xi_1)\cdots\cL_{A,2n}(\z/\xi_{2n})\right)
\Bigr)
i^{-2(s+1)\overline{S}}\,.
\en
The assertion follows from Lemma \ref{lem:Qt} and the relation  
\be
i^{2(s+1)\overline{S}}
\theta(\phi^{*\pm}_j(\al)\phi^\mp_k)
i^{-2(s+1)\overline{S}}
\bigl|_{\iota(\cV^{(s)})}
=\phi^\mp_k\phi^{*\pm}_j(\al)
\bigl|_{\iota(\cV^{(s)})}
\en

\end{proof}
\medskip

\noindent{\it Proof of Theorem \ref{thm:tQ}}.\quad 
We apply Lemma \ref{lem:Qt} to Lemma \ref{lem:QQ*}, 
replacing $\al$ by $\al-s$ and noting that  
$\phi^{*\pm}_j(\al)\phi^\mp_k\iota(X)=
\iota\bigl(\Phi^\pm_j(\al+s)\Psi^\mp_k(X)\bigr)$
for $X\in\cV^{(s)}$. 
We find 
\be
&&\z^{-\al+\bS}\bQ^+(\z,\al)
\\
&&=
\exp
\left(\sum_{j,k=1}^n
\bigl(\log(1+\z^2 M)_{jk}\Phi^{+}_j(\al+1)\Psi^-_k
 -\log(1-\z^2 M)_{jk}\Phi^-_j(\al+1)\Psi^+_k\bigr)\right). 
\en
This expression 
can be further simplified by diagonalizing 
the matrix $M$ as
\be
PMP^{-1}={\rm diag}(\xi_1^{-2},\cdots,\xi_n^{-2}), 
\en
where $P$ is the matrix given in \eqref{eq:P}. 
We thus obtain the result
\be
&&\z^{-\al+\bS}\bQ^+(\z,\al)
\\
&&=
\exp
\left(\sum_{k=1}^n
\bigl(\log(1+\z^2/\xi^2_{k})\Phih^+_k(\al+1)\Psih^-_k
 -\log(1-\z^2/\xi^2_{k})\Phih^-_k(\al+1)\Psih^+_k
\bigr)
\right), 
\en
which is equivalent to \eqref{fQ}. 

Similarly, \eqref{fQ*} can be shown by Lemma \ref{lem:QQ*} and 
Lemma \ref{lem:Qt*}. 

{}Finally, formulas \eqref{ft},\eqref{ft*}
follow from \eqref{fQ}, \eqref{fQ*} and 
the TQ relations \eqref{TQ}, \eqref{TQ*}.
\qed

\subsection{A factorization}

Let us proceed to the calculation of $\bb(\z),\cb(\z)$. 
A simplifying feature about the free fermion point is a 
factorization property. 

\begin{prop}\label{prop:factor}
The following factorization takes place:
\bea
&&
\bk^\pm(\z,\al)=
\Psih^\pm(\z)\bQ^\pm(-i\z,\al\mp1)i^{\pm \al}\,,
\label{eq:factor1}\\
&&
\bk^{*\pm}(\z,\al)
=-i\Phih^\pm(i\z)
(i^{\al-\bS}+i^{-\al+\bS})
\bQ^{\pm}(\z,\al)\,.
\, 
\label{eq:factor2}
\ena
\end{prop}
\begin{proof}
{}First consider the action of $\bk^+(\z,\al)$ on $X\in \cV^{(s)}$. 
Inserting \eqref{TRIG} in the definition \eqref{fusion} 
and taking the trace over $V_a$, we obtain $2n$ terms: 
\be
&&
{\rm tr}_a\bigl(\sigma^+_a
{T_{\{A,a\}}^+(\z)}^{-1}\ X\ T_{\{A,a\}}^+(\z)\bigr)
\\
&&=
-\sum_{j=1}^n\gamma(\z/i\xi_j)\prod_{l=1}^{j-1}\beta(\z/\xi_l)^{-1}
i^{-\sigma_1^3/2}L^+_{A,1}(i\z/\xi_1)^{-1}
\cdots \sigma^+_ji^{-2D_A-1/2}L^+_{A,j}(\z/i\xi_j)^{-1}
\\
&&\quad 
\times 
\cdots i^{\sigma^3_n/2}L^+_{A,n}(\z/i\xi_n)^{-1}
\cdot X
L^+_{A,n}(\z/i\xi_n)i^{-\sigma^3_n/2}
\cdots
L^+_{A,1}(\z/i\xi_1)i^{-\sigma^3_1/2}
\\
&&+
\sum_{j=1}^n\gamma(\z/\xi_j)\prod_{l=1}^{j}\beta(\z/\xi_l)^{-1}
i^{-\sigma_1^3/2}L^+_{A,1}(i\z/\xi_1)^{-1}
\cdots i^{-\sigma^3_n/2}L^+_{A,n}(i\z/\xi_n)^{-1}
\\
&&\quad 
\times X\,
L^+_{A,n}(i\z/\xi_n)i^{-\sigma^3_n/2}
\cdots
i^{-\sigma_j^3}L_{A,j}^+(\z/i\xi_j)\sigma_j^+i^{-2D_A+1/2}
\cdots
L^+_{A,1}(\z/i\xi_1)i^{-\sigma^3_1/2},
\en
where in the second sum we have used 
\be
L^+_{A,j}(i\z/\xi_j)\sigma_j^+
=i^{-\sigma_j^3}L_{A,j}^+(\z/i\xi_j)\sigma_j^+\,.
\en
Rewriting this expression in terms of fermions, we obtain 
\be
&&
\iota\Bigl({\rm tr}_a\bigl(\sigma^+_a
{T_{\{A,a\}}^+(\z)}^{-1}\ X\ T_{\{A,a\}}^+(\z)\bigr)\Bigr)
\\
&&=
\sum_{j=1}^n\frac{2i \z/\xi_j}{1-\z^2/\xi_j^2}
\frac{1}{\prod_{l=1}^{j-1}(1-\z^2/\xi_l^2)
\prod_{l=j}^{n}(1+\z^2/\xi_l^2)}
\cdot i^{-(2D+1)\widetilde{S}}
\\
&&\times
\bigl(
\cL_{A,1}(i\z/\xi_1)
\cdots
\cL_{A,j-1}(i\z/\xi_{j-1})
\psit^+_j\cL_{A,j}(\z/i\xi_j)
\cdots 
\cL_{A,n}(\z/i\xi_n)\cL_{A,\bar{n}}(\z/\xi_n)
\cdots
\cL_{A,\bar{1}}(\z/\xi_1)
\\
&&+
i\cL_{A,1}(i\z/\xi_1)\cdots \cL_{A,n}(i\z/\xi_n)
\cL_{A,\bar{n}}(-\z/\xi_n)\cdots
\psit^+_{\bar{j}}
\cL_{A,\bar{j}}(\z/\xi_j)
\cdots 
\cL_{A,\bar{1}}(\z/\xi_1)\bigr)
\cdot\iota(X)\,.
\en

Since 
\be
\cL_{A,k}(\z)\psit^\pm_j=\psit^\pm_j\cL_{A,k}(-\z)
\quad (j\neq k)\,, 
\en
we can bring $\psit^+_j$ to the leftmost place. 
The result factorizes into a linear form in the fermion
\bea
\sum_{j=1}^n\frac{2\z/\xi_j}{1-\z^2/\xi_j^2}
\prod_{l=1}^{j-1}\frac{1+\z^2/\xi_l^2}{1-\z^2/\xi_l^2}
\left(i\psit^+_j-\psit^+_{\bar{j}}\right)
\label{fpart}
\ena
followed by the product 
\bea
\prod_{l=1}^n\frac{1}{1+\z^2/\xi_j^2}\cdot
i^{-(2D_A+1)(\widetilde{S}+1)}
\cL_{A,1}(-i\z/\xi_1)
\cdots \cL_{A,2n}(-i\z/\xi_{2n})\cdot\iota(X)\,.
\label{Qpart}
\ena
Multiply both sides by 
$\z^{\al-\bS}(1-i^{2(\al-\bS)})i^{2\al D_A}$  and take the 
trace $\Tr^+_A$. 
In view of the relations
\be
&&
i\psit^\pm_j\cdot\iota(X)=(-1)^{s+1}\cdot
\iota\left(\psi^\pm_j\ X\right),
\quad 
 -\psit^\pm_{\bar{j}}\cdot\iota(X)=
\iota\left(X \psi^\pm_j\right)\,,
\en
the piece \eqref{fpart} yields $(-1)^{s+1}\Psih^+(\z)$. 
On the other hand, due to \eqref{bQ1}, 
trace of \eqref{Qpart} (taken together with the prefactor) 
gives rise to $(-1)^{s+1}i^{\al}\bQ^+(-i\z,\al-1)(X)$. 
Combining these we obtain \eqref{eq:factor1}. 

Similarly, using
\be
\bigl(
i^{s-\al}\psit^-_j-i^{-s+\al+1}\psit^-_{\bar{j}}\bigr)\iota(X)
=(i^{-\al-s+1}-i^{\al+s-1})\iota\bigl(\Phi^-_j(\al)(X)\bigr)\,,
\en
we compute
\be
&&
\bk^{*-}(\z,\al)(X)
=-i\Phih^-(i\z,\al)\, (1+i^{2(\al-\bS)})
\prod_{l=1}^n\frac{1+\z^2/\xi_l^2}{1-\z^2/\xi_l^2}
\cdot i^{-2S}\bQ^{*-}(i\z,\al)(i^{2S}X)\,.
\en
On the other hand, \eqref{fQ} and \eqref{fQ*} imply
\be
\prod_{l=1}^n\frac{1+\z^2/\xi_l^2}{1-\z^2/\xi_l^2}
\cdot i^{-2S}\bQ^{*-}(i\z,\al)(i^{2S}X)
=
i^{-\al+\bS}\bQ^{-}(\z,\al)(X). 
\en
The assertion follows from these.  
\end{proof}

\medskip

\noindent{\it Proof of Theorem \ref{thm:c}}.
\quad 
{}From Proposition \ref{prop:factor} and Theorem  \ref{thm:tQ},
it is simple to calculate the residues of $\bk^\pm(\z,\al)$. 
We have 
\be
{\rm res}_{\z=\xi_k}\bk^\pm(\z,\al)\frac{d\z}{\z}
=
-\frac{1}{2}\Psih^\pm_k\cdot U^\pm_k(\al)(1-N^\pm_k(\al))
\cdot
i^{\pm\bS+1}\xi_k^{\pm(\al-\bS)-1}.
\en
After simplification using $1-N^\pm_k=\chi^\mp_k\chi^{*\pm }_k$, 
we obtain the desired expression for $\cb^\pm_k$.  
\qed
\bigskip

\noindent
{\it Acknowledgments.}\quad
Research of HB is supported by the RFFI grant \#04-01-00352.
Research of MJ is supported by 
the Grant-in-Aid for Scientific Research B--18340035.
Research of TM is supported by 
the Grant-in-Aid for Scientific Research B--17340038.
Research of FS is supported by INTAS grant \#03-51-3350, by EC networks  "EUCLID",
contract number HPRN-CT-2002-00325 and "ENIGMA",
contract number MRTN-CT-2004-5652
and GIMP program (ANR), contract number ANR-05-BLAN-0029-01. 
Research of YT is supported by the Grant-in-Aid for Young Scientists B--17740089. 
This work was also supported by the grant of 21st Century 
COE Program at RIMS, Kyoto University. 

MJ would like to thank 
L. Takhtajan and J.M. Maillet for the interests and discussions. 
MJ and TM are grateful to 
LPTHE Universit{\'e} Paris VI for kind hospitality, 
where a part of this work has been carried out.


\bigskip


\begin{thebibliography}{[FJKLM]}

\bibitem{BJMST} 
H.~Boos, M.~Jimbo, T.~Miwa, F.~Smirnov and
Y.~Takeyama,  
\newblock Hidden Grassmann structure in the XXZ model, 
\newblock hep-th/0606280, 
to appear in {\em Commun. Math. Phys. }

\bibitem{BLZ}
V.~Bazhanov, S.~Lukyanov and A.~Zamolodchikov,
\newblock 
Integrable structure of conformal field theory III.
The Yang-Baxter relation, 
\newblock {\em Commun. Math. Phys. }{\bf 200}
(1999) 297--324.
\end{thebibliography}
\end{document}